\documentclass[twocolumn]{aastex631}

\providecommand{\e}[1]{\ensuremath{\times 10^{#1}}}

\begin{document}

\title{Detectability of Emission from Exoplanet Outflows Calculated by pyTPCI, a New 1D Radiation-Hydrodynamic Code}


\correspondingauthor{Riley Rosener}
\email{rosener.riley@gmail.com}

\author[0000-0001-7905-2134]{Riley Rosener}
\affiliation{Department of Astronomy \& Astrophysics \\
University of Chicago \\
5640 S Ellis Ave \\
Chicago, IL 60637, USA}

\author[0000-0002-0659-1783]{Michael Zhang}
\affiliation{Department of Astronomy \& Astrophysics \\
University of Chicago \\
5640 S Ellis Ave \\
Chicago, IL 60637, USA}

\author[0000-0003-4733-6532]{Jacob L. Bean}
\affiliation{Department of Astronomy \& Astrophysics \\
University of Chicago \\
5640 S Ellis Ave \\
Chicago, IL 60637, USA}


\begin{abstract}
Photoevaporation in exoplanet atmospheres is thought to contribute to the shaping of the small planet radius valley. Escaping atmospheres have been detected in transmission across a variety of exoplanet types, from hot Jupiters to mini-Neptunes. However, no work has yet considered whether outflows might also be detectable in emission.  We introduce pyTPCI, a new, open-source self-consistent 1D radiative-hydrodynamics code that is an improved version of The PLUTO-CLOUDY Interface. We use pyTPCI to model seven exoplanets (HD 189733b, HD 209458b, WASP-69b, WASP-107b, TOI-1430b, TOI-560b, and HAT-P-32b) at varying metallicities and compute their emission spectra to investigate their detectability across a variety of spectral lines. We calculate the eclipse depths and signal-to-noise ratios (SNR) of these lines for a 10m class telescope with a high-resolution spectrograph, taking into account appropriate line broadening mechanisms.  We show that the most detectable spectral lines tend to be the 589 nm Na I doublet and the 1083 nm metastable helium triplet.  H$\alpha$ and Mg I 457 nm are moderately strong for some planets at some metallicities, but they are almost always optically thin, so some of their emission may not be from the outflow. The planet with the highest-flux, highest-eclipse-depth, and highest-SNR lines is HD 189733b, with a Na I eclipse depth of 410 ppm and SNR of 2.4 per eclipse, and a He$^*$ eclipse depth of 170 ppm and SNR of 1.3. These signals would be marginally detectable with Keck if 3--10 eclipses were observed, assuming (over-optimistically) photon limited observations.
\end{abstract}



\section{Introduction} \label{sec:intro}
The study of escaping exoplanet atmospheres is only 20 years old \citep{VidalMadjar03_hd209}, and yet the development of novel approaches to atmospheric modeling and observing is still rapidly advancing the field. Exoplanets close to their host stars experiencing strong irradiation at high-energy wavelengths will gradually have their atmospheres stripped away via photoevaporation. Although the specific mechanisms and timescales of this process remain controversial and not yet fully understood, it is vital for understanding exoplanet demographics, particularly the radius gap between super-Earths and mini-Neptunes that it likely has a key role in sculpting \citep{Fulton17_radiusgap, Fulton18_keplervii, Owen13_kepler, Owen23_cpmassloss}. 

The first outflow tracer to be observed was the Ly-$\alpha$ line, which traces neutral hydrogen. Escaping neutral hydrogen was first observed around the archetypal hot Jupiter HD 209458b via absorption of stellar Lyman-$\alpha$ \citep{VidalMadjar03_hd209}.
For this planet, \citet{MurrayClay09_escape} argue that photoevaporation cannot strip a significant fraction of mass via atmospheric outflows. However, the Ly-$\alpha$ line core is subject to absorption by the interstellar medium and contamination by geocoronal emissions, making it difficult to observe. 
 Nonetheless, Ly-$\alpha$ absorption was later detected in hot Jupiter HD 189733b, constraining the atomic hydrogen mass loss rate to be a similar value \citep{Lecavelier10_hd189}, as well as in a few smaller planets: the warm Neptune GJ 436b \citep{Ehrenreich15_gj436}, the mini-Neptune HD 63433c \citep{Zhang22_hd63433}, and (tentatively) the mini-Neptune K2-18 b \citep{DosSantos20_k218}.

Building off earlier Ly-$\alpha$ detections, \citet{Jensen12_halpha} identified H$\alpha$ absorption in the outflows of HD 189733b and HD 209458b, which were used to estimate the hydrogen excitation temperature \citep{Christie13_Halpha}. After detecting H$\alpha$ and the Na I doublet in transit, \citet{CasasayasBarris18_kelthalpha} recommended further measurements of KELT-20b in H$\alpha$ and the Balmer lines to map the ultra-hot Jupiter's temperature profile and estimate its mass loss rate. 

\citet{Seager00_hd209458} first proposed the metastable helium (He$^*$) triplet at 10833\, \AA\ as a promising tracer of the exosphere. Following the analysis of \citet{Oklopcic18_He}, metastable helium absorption has been detected in various exoplanet outflows, such as in the Saturnian exoplanets WASP-107b \citep{Spake2018_wasp107} and WASP-69b \citep{Nortmann18_wasp69}, the hot Jupiter HD 189733b \citep{Salz18_hd189733}, and various mini-Neptunes such as TOI-560b \citep{Zhang22_toi560}, TOI-2134b \citep{Zhang23_toi2134}, TOI-1430b \citep{Zhang23_neptunes}, TOI-2018b \citep{OrellMiquel24_MOPYS_preprint}, and TOI-836c \citep{Zhang24_toi836}. However, mass loss estimates for HD 189733b alone vary over multiple orders of magnitude from $4 \times 10^9 \mathrm{g} \, \mathrm{s}^{-1}$ \citep{Salz16_escaping} to $\sim 1.1 \times 10^{11} \mathrm{g} \, \mathrm{s}^{-1}$ \citep{DosSantos23_hd189733}, indicating significant differences in methodology and modeling that must be examined. No clear statistical analysis of what precisely drives helium absorption detection and non-detection has yet emerged, though some preliminary demographic surveys have been undertaken \citep{Guilluy24_GAPS_He, OrellMiquel24_MOPYS_preprint}.

HD 189733b's escaping atmosphere was modeled \citep{Salz18_hd189733} using TPCI, The PLUTO-CLOUDY Interface \citep{Salz15_TPCI, Salz16_escaping}. TPCI is a 1D radiative-hydrodynamics code that couples the hydrodynamics code PLUTO and the gas microphysics code CLOUDY to simulate mass loss in a self-consistent way. TPCI benefits in drawing from the well-tested and widely used code bases of CLOUDY and PLUTO, though at the cost of computational speed, since neither code was optimized for modeling mass loss.

Other escaping atmospheres codes exist: \citet{Caldiroli21_ATES} developed the 1D ATES photoionization hydrodynamics code. This code agrees with the TPCI results of \citet{Salz16_escaping} to a factor of two while running much more quickly. However, it does not take into account ion advection during the simulation, does not include metals, conductivity, and radial mixing of different species, and fails to converge for the heaviest gas giants that were modeled. \citet{DosSantos22_pwinds} presented \verb|p-winds| to retrieve the hydrogen and helium mass loss rate in the simple 1D isothermal Parker wind forward model proposed by \citet{Oklopcic19_He} and \citet{Lampon20_hd209458}. \citet{Linssen24_sunbather} combined \verb|p-winds| and CLOUDY to create \verb|sunbather|, which relaxes the isothermal assumption, but is still not self-consistent because the mass loss rate is a free parameter \citep{Linssen22_cloudy}. CHAIN \citep{kubyshkina_2023} is self-consistent, coupling a custom hydrodynamic solver with CLOUDY, but it is not publicly available.  Until this work, TPCI was the only publicly available code that self-consistently models outflows while including metal heating and cooling, but despite that it had not seen widespread adoption, perhaps because it is no longer maintained. Building off of TPCI, we introduce pyTPCI, a Python code for coupling PLUTO and CLOUDY together to model atmosphere outflows. It uses up-to-date PLUTO and CLOUDY releases, includes various fixes to improve stability and speed, adds logging capabilities, and features easy-to-use template files. 

An extensive literature exists on characterizing exoplanet outflows via transmission spectroscopy, but few published works have  considered outflows in emission (e.g. \citet{ZhangSnellen20_He}). Exoplanet outflows could be hot and energetic enough to be emitting significant flux in many lines. In this paper, we explore that possibility. These emissions would open a new window into escaping atmospheres. They could provide valuable information about the temperature profile, density, ionization state, and metallicity of the hot outflow, in addition to allowing us to test and refine models of atmospheric escape. 

In this work, we calculate the observability of signals from outflow emissions using the updated and improved radiative-hydrodynamics simulation pyTPCI to model the outflow. We examine the feasibility of observing the brightest lines of the simulated emission spectra.

We describe the features and development of pyTPCI in Section \ref{sec:pytpci} and the simulation setup, parameters, and choice of our grid of planets in Section \ref{sec:setup}. We explain our calculations of detectability measures in Section \ref{sec:results} and discuss the implications, observational potential, and limitations of these results in Section \ref{sec:discussion}, concluding in Section \ref{sec:conclusion}. We also briefly examine the effects of different stellar spectra in Appendix \ref{sec:apptab}.

\section{pyTPCI} \label{sec:pytpci}
\cite{Salz15_TPCI} introduced The PLUTO-CLOUDY Interface (TPCI): a 1D radiation-hydrodynamic code that couples the hydrodynamic code PLUTO \citep{Mignone07_PLUTO} to a lightly modified version of the spectral synthesis code CLOUDY \citep{ferland_2013} to self-consistently simulate outflows. Both PLUTO and CLOUDY are mature and sophisticated codes that are widely used across many fields of astrophysics. In TPCI, PLUTO evolves the fluid's density, pressure, and velocity based on heating and cooling rates computed by CLOUDY.  CLOUDY computes heating and cooling rates based on the density, pressure, and velocity profile from PLUTO and a user-specified stellar spectrum. Initially conceived for the interstellar medium, physical processes that CLOUDY takes into account include photoionization/recombination, collisional excitation/de-excitation, atomic line cooling, and advection.

TPCI was the most self-consistent open source outflow simulation code that we are aware of, but it is no longer being maintained.  While powerful and unique, it did have a few weaknesses--it took a long time to run (\citealt{Salz16_escaping} reports 300,000 CPU hours for their 14 planets), and because it is no longer maintained, the versions of PLUTO and CLOUDY that it uses are out of date and therefore hard to install.  The programming language (C++) may also pose difficulties to users more familiar with Python.  To improve upon TPCI, we release pyTPCI, a Python wrapper which couples much newer versions of PLUTO and Cloudy (4.4-patch1 and C23 respectively, released in 2020 and 2023). Following TPCI, we modified CLOUDY to read in depth-dependent wind velocities, and to output useful physical quantities (including mass density, number density, and mean molecular weight as a function of depth). pyTPCI is very easy to install, and in most cases converges within 1--2 days when run on a Core i9 13900k CPU, although high metallicities are still prone to numerical instabilities that cause the code to crash or take very small timesteps.

pyTPCI differs from TPCI in several other ways. First, when we apply CLOUDY's computed volumetric heating rate in PLUTO, we scale the heating rate by the instantaneous density (essentially turning it into a heating rate per unit mass). This makes the code much more stable because it prevents a runaway cycle whereby heating the gas decreases its density, which increases the heating rate per unit mass, and so on until the simulation crashes. With this change, we can take large timesteps in the initial relaxation phase when the stellar irradiation is first turned on and a heating wave propagates outward through the simulation domain. Second, instead of simulating the substellar point, we simulate a point more representative of the global average by adopting a 66$^\circ$ zenith angle for the incident stellar irradiation. We adopt this angle because \cite{johnstone_2018} found that for Earth's upper atmosphere, it gives a decent approximation of globally averaged profiles. Third, we make the radial coordinate grid twice as dense at lower radii due to the extremely high density and pressure gradients there, and expand the grid from 15 $R_p$ to 30 $R_p$ so that the outflow from smaller planets can cover the stellar disk. Finally, we add a small amount of numerical shear viscosity $\nu$ to PLUTO via super-time-stepping in an attempt to improve numerical stability. The numerical viscosity gives a Reynolds number $\frac{\Delta U}{\nu} \sim \mathcal{O}(1)$ at a flow speed $U$ of 10 km/s and characteristic length $\Delta$ of 0.0002 $R_p$, the smallest grid size in our simulation.

pyTPCI is open source and available on GitHub at \url{https://github.com/ideasrule/pyTPCI}. We encourage users to apply it to a wide variety of outflows and explore the code's strength and limitations.

\begin{deluxetable*}{ccccccccccc}[ht]

\tablecaption{pyTPCI Input Parameters \label{tab:params}}

\tablehead{\colhead{System} & \colhead{$T_{\rm eff,*}$} & \colhead{$F_{\rm XUV}$} & \colhead{Roche Radius} & \colhead{Coriolis Radius} & \colhead{Distance} & \colhead{Semimajor Axis} & \colhead{Mass} & \colhead{Radius} & \colhead{T$_{\rm eq}$} & \colhead{Metallicity} \\ 
\colhead{} & \colhead{(K)} & \colhead{(ergs s$^{-1}$ cm$^{-2}$)} & \colhead{($R_{\rm RL}/R_{*}$)} & \colhead{($R_{\rm C}/R_{*}$)} & \colhead{(pc)} & \colhead{(AU)} & \colhead{($M_\earth$)} & \colhead{($R_\earth$)} & \colhead{(K)} & \colhead{($\times$ solar)} } 

\startdata
HD 189733b & 5012 & 18 & 0.458 & 1.956 & 19.76 & 0.03126 & 370.6 & 12.54 & 1209 & 0, 1, 10 \\
HD 209458b & 6026 & 3.9 & 0.336 & 1.556 & 48.3 & 0.04723 & 219 & 15.1 & 1320 & 0, 1 \\
WASP-69b & 4715 & 6.8 & 0.390 & 2.424 & 49.96 & 0.045 & 92 & 12.4 & 963 & 0, 1 \\
WASP-107b & 4400 & 8.0 & 0.473 & 2.687 & 64.74 & 0.0553 & 37.8 & 10.4 & 770 & 0, 1, 10 \\
TOI-560b & 4500 & 3.5 & 0.340 & 0.966 & 31.6 & 0.0596 & 11 & 2.9 & 740 & 0, 1, 10, 100 \\
TOI-1430b & 5037 & 9.5 & 0.280 & 0.925 & 41.17 & 0.072 & 7 & 2.04 & 800 & 0, 1, 10, 100 \\
HAT-P-32b & 6000 & 350 & 0.215 & 0.309 & 289.21 & 0.03397 & 216 & 22.19 & 1835 & 0, 1 \\
\enddata

\tablecomments{Fluxes given at 1 au from host system. XUV flux spans 1--504 \AA. Stellar temperature, distance, semimajor axis, radius, and equilibrium temperature are taken from the NASA Exoplanet Archive.}
\tablerefs{Planet masses. HD 189733b: \citet{Addison19_hd189733}; HD 209458b: \citet{Bonomo17_hd209458}; WASP-69b: \citet{Stassun17_wasp69}; WASP-107b: \citet{Mocnik17_wasp107}; TOI-560b: \citet{Barragan21_toi560}; TOI-1430b (via mass-radius relation): \citet{Wolfgang16_subneptune}; HAT-P-32b: \citet{Wang19_hatp32}.}
\end{deluxetable*}

\section{Simulation Setup} \label{sec:setup}
In this section, we describe how we set up the initial conditions of pyTPCI, and note some computational considerations for running it.
The operational parameters of pyTPCI were chosen via experimentation to optimize simulation stability. When initializing pyTPCI, one must specify the temperature and the number density of the gas at all radii. The temperature profile is set to be a uniform distribution at the planet's equilibrium temperature. The mean molecular weight and hydrogen fraction of the gas are determined by the assumed metallicity. The surface pressure is determined using the ideal gas law and initial temperature, assuming the number density at the bottom of the simulation domain is $10^{14}\,\mathrm{g}\,\mathrm{cm}^{-3}$. The initial atmospheric pressure profile is calculated from hydrostatic equilibrium. The initial number density profile in PLUTO is computed from the pressure and temperature.  

The construction of our grid---the combinations of planets and metallicities---was intended to explore a wide parameter space within which escaping metastable helium has already been detected. Thus, our systems consist of two hot Jupiters (HD 189733b and HD 209458b), two Saturnian exoplanets (WASP-69b and WASP-107b), two mini-Neptunes (TOI-560b and TOI-1430b), and one ultra-hot Jupiter (HAT-P-32b). The systems also cover a diversity of stellar spectral types: the majority are K dwarfs, but HD 209458 is a G dwarf and HAT-P-32 is a F dwarf.

For a given system, the planet parameters such as planet mass, radius, and semimajor axis were taken from the most recent literature. Equilibrium temperatures were taken from the NASA Exoplanet Archive (\url{https://exoplanetarchive.ipac.caltech.edu/}), where sources vary on their assumptions and modeling methodologies. Generally, simulations for each system were run for metallicities of 0$\times$, 1$\times$, 10$\times$, and 100$\times$ solar.  However, higher metallicity generally results in worse stability in CLOUDY, and sometimes 100$\times$ solar metallicity runs were not able to be run to completion. Runs with non-zero metallicity include elements with solar abundances greater than 10$^{-5}$: H, He, O, C, Ne, N, Si, Mg, Fe, and S \citep{Chatzikos23_CL23}. We additionally include K on the final step CLOUDY is run.

X-rays and extreme UV (XUV) radiation is predominantly responsible for driving the outflow through photoionization heating \citep{Oklopcic18_He}. For each of our systems, we use stellar XUV spectra (1--912\,\AA) from Emission Measure Distribution (EMD) models of the stellar corona and chromosphere, as described in \citet{SanzForcada25_He}. For the wavelength range between 912 and 50,000\,\AA, we also add contributions from the stellar photosphere from the appropriate PHOENIX high-resolution synthetic stellar spectra \citep{Husser13_PHOENIX}. Appendix \ref{sec:apptab} tabulates alternative results using the stellar spectrum construction method of \citet{Salz16_escaping}.

In pyTPCI, we define one time unit to be the approximate sound crossing time: the planet radius $R_p$ divided by a typical sound speed $U=10$ km/s. In most cases, we initially run pyTPCI with advection turned off until $t=100$ for a timestep of $dt=0.01$, and then run it with advection turned on until $t=1,000$.  We then manually check the simulation to ensure it is converged. When the simulation starts and stellar irradiation first turns on, a large temperature spike forms near the base of the simulation and propagates outward.  If advection is turned on at this stage, the spike grows extremely large and crashes CLOUDY. For less stable cases like HD 189733b, we must use a smaller timestep $dt=0.001$. In order to have the simulation finish reasonably quickly, once the density profile does not change much between timesteps we start advection earlier at $t=5$, then run the simulation as long as feasible until the results are stable around $t=55$.

A notable phenomenon exhibited by both TPCI and pyTPCI is the tendency for stable sound waves to form, particularly in simulations of HD 189733b. These are artifacts resulting from small numerical errors building up in the simulation. We were able to measure the wave period and speed, and confirm that the waves were traveling at the speed of sound. By fixing the time step to the wave's period and restarting pyTPCI from where it crashed, we successfully froze the waves in place and could thereby run the simulations with larger timesteps that were multiples of the period.


For certain parameter configurations, pyTPCI is inherently unstable, driving the time step to be so low that PLUTO crashes very quickly, usually on the first few time-steps near $t=0.10$. This occurs while using PLUTO's ``Super-Time-Stepping (STS)'' numerical integration method for including the viscosity and thermal conduction terms, which runs much more quickly than explicit time-stepping by taking sub-steps in time. When this error occurs, the length of the sub-timestep calculated to be necessary for a given grid size and diffusion Courant number of $0.8/N_{\rm dim}$ drops below PLUTO's minimum allowed timestep. This is most common for very large and hot planets simulated at nonzero metallicity, such as HD 189733b at 100$\times$ solar, HD 209458b at 10$\times$ solar, and HAT-P-32b at 10$\times$ solar. Those models were unable to run to completion, and thus that section of the parameter space remains unexplored. 

A successful pyTPCI simulation takes approximately 48 hours wall time to converge on an Intel Core i9 13900k processor with 64 GB RAM.  There is significant variation in run time because of differing convergence behavior. For example, HD 189733b at 10$\times$ solar metallicity takes around 240 hours of wall time.

\section{Calculations of Observed Signals and Results} \label{sec:results}

We analyze the pyTPCI output files in order to calculate the eclipse depth and signal-to-noise ratio (SNR) of various spectral lines from the outflow. We identify lines present in the spectrum, and assess the viability of detecting any significant signals from our targets using 10m class telescopes equipped with high-resolution spectrographs with a total throughput of 10\%. 
Some specific instruments fit these parameters, namely the Keck Planet Finder (KPF) \citep{Gibson24_KPF} and the Near Infrared Spectrometer (NIRSPEC) \citep{McLean98_NIRSPEC} on Keck. KPF is an example of a modern stabilized high-resolution optical spectrograph. NIRSPEC is an instrument with a proven track record in detecting escaping helium, and holds the record for the smallest detected helium signal: 0.4\% excess absorption from the mini-Neptune TOI-2134b \citep{Zhang23_toi2134}.

For KPF, we assume a 10\% throughput with a resolution of 98,000.\footnote{\url{https://www2.keck.hawaii.edu/inst/kpf/kpf_vs_hires.html}} For NIRSPEC, the seeing-limited Y-band filter with the 0.720$\times$12'' slit has a throughput of 19.5\%\footnote{\url{https://www2.keck.hawaii.edu/inst/nirspec/sens.html}}, but since observations such as \citet{Zhang22_toi560} use the 0.432$\times$12'' slit, we adopt a smaller throughput of 10\% for simplicity. We also assume a resolution of 32,000.

CLOUDY outputs the emission spectrum in a low-resolution (R=300) \verb|continuum.tab| file. CLOUDY can be run at much higher resolution, but it was not designed for high resolution (M. Chatzikos, private communication) and can encounter numerical issues. Thus, we approximate each emission line with a Gaussian so that it has an integrated flux equal to that of the line in the pyTPCI spectrum, but with a realistic width. This allows us to simulate what a high-resolution spectrograph would observe at the line core without running CLOUDY at resolutions of $\gtrsim20,000$.

In order to compute the spectral line width, we take into account the effects of thermal broadening at an outflow temperature of $T_{\rm outflow}$, wind broadening from the velocity of the outflow $v_{\rm outflow}$, rotational broadening assuming all planets are tidally-locked with an equatorial velocity of $v_{\rm rot}$, and broadening as a result of instrumental resolution $R_{\rm instr}$, as shown in Equation \ref{eq:width}. This computes the spectral line width at wavelength $\lambda$ for an outflow with a mean molecular weight $\mu$.
\begin{equation}\label{eq:width}
    \sigma_{\lambda} =  \frac{\lambda}{c} \cdot \sqrt{\frac{k_B T_{\mathrm{outflow}}}{\mu m_H} + v_{\rm outflow}^2 + v_{\rm rot}^2 + \Big(\frac{c/R_{\rm instr}}{2\sqrt{2\ln2}}\Big)^2} 
\end{equation}

CLOUDY reports flux per unit of emitting area (spectral radiance). The emitting area could be simply $4\pi R_p^2$ if the line is optically thin, or it could be larger if the photosphere is higher in the outflow.
In order to calculate the photosphere radius, we calculate where each line becomes optically thick ($\tau=1$). For each line, we compute the absorption cross section $\sigma$ as a function of the line oscillator strength $f_s$, taking the line profile $\phi_{\nu}$ to be a Gaussian with a width given by Equation \ref{eq:width} minus the instrumental broadening \citep[eq. 6.24]{Draine11_IGM}:
\begin{equation}\label{eq:crosssect}
    \sigma = \frac{\pi e^2}{m_e c}f_s \cdot \phi_{\nu} 
\end{equation}

Integrating from the top to the bottom of the outflow and using the depth-dependent number densities from CLOUDY output files, we find the optical depth as a function of the radial coordinate for each line. If the optical depth never exceeds 1, the line is always optically thin, and the photosphere radius is taken to be the white light radius. In such cases, some of the line emission may come not from the outflow, but from the lower atmosphere, which we do not model.

Using the photosphere radius, we find the maximum planet-to-star flux ratio at the peak of the Gaussian observed by the high-resolution spectrograph. We also find the narrowband eclipse depth within $\pm 2w$ of the line center, where $w$ is the standard deviation of the Gaussian. In both cases, we multiply the ratio of planetary to stellar surface fluxes by the square of photosphere-radius-to-stellar-radius ratio:

\begin{equation}\label{eq:ED}
    \delta(\lambda) = \frac{F_{p,\lambda}}{F_{*,\lambda}} \cdot \Big(\frac{R_{\mathrm{photosphere}}}{R_{*}}\Big)^2
\end{equation}

\begin{figure}
	\centering
	\includegraphics[width=0.5\textwidth]{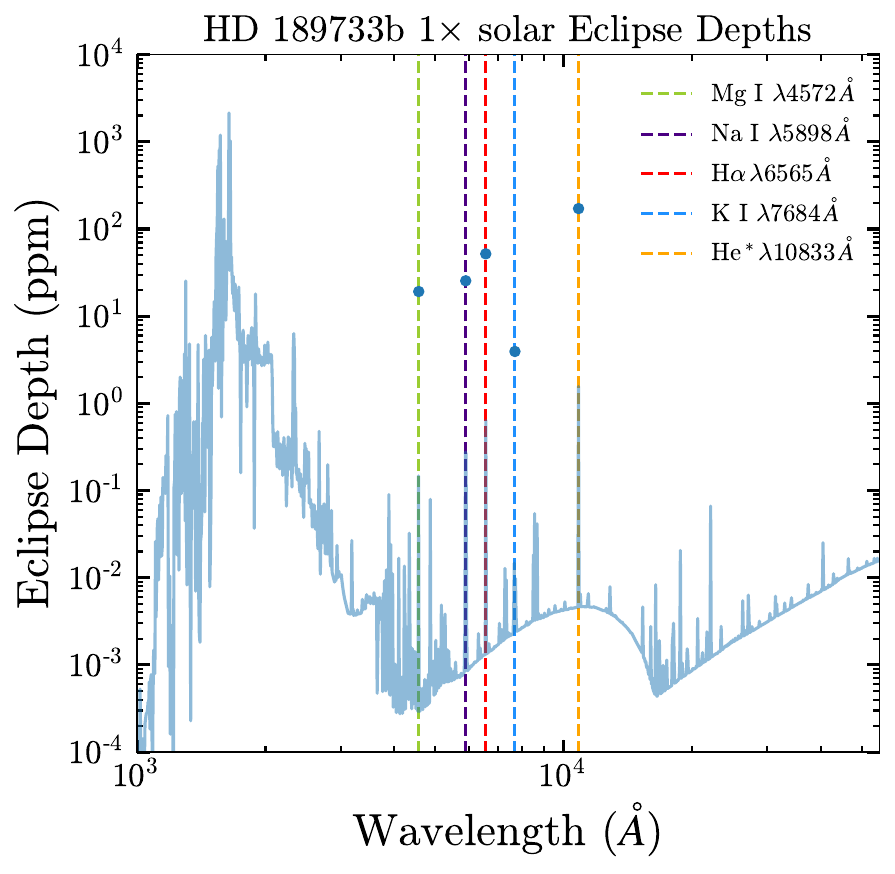}
	\caption{Dots for certain lines are the maximum eclipse depths taking into account broadening and resolution effects. The line is the R=300 eclipse depths.}
	\label{fig:hd189_1Z_ED}
\end{figure}

The narrowband eclipse depth, together with the stellar spectrum, the stellar distance, the total throughput, and the observation duration allow us to calculate the SNR to which the line can be detected in one eclipse observation. We take the observation duration to be twice the planet's eclipse duration, covering the eclipse and an equal amount of baseline. The maximum eclipse depths are plotted as dots in Figure \ref{fig:hd189_1Z_ED}, with the background line being the low-resolution (R=300) eclipse depths calculated from the low-resolution emission spectrum that CLOUDY outputs.

\begin{figure}
	\centering
	\includegraphics[width=0.5\textwidth]{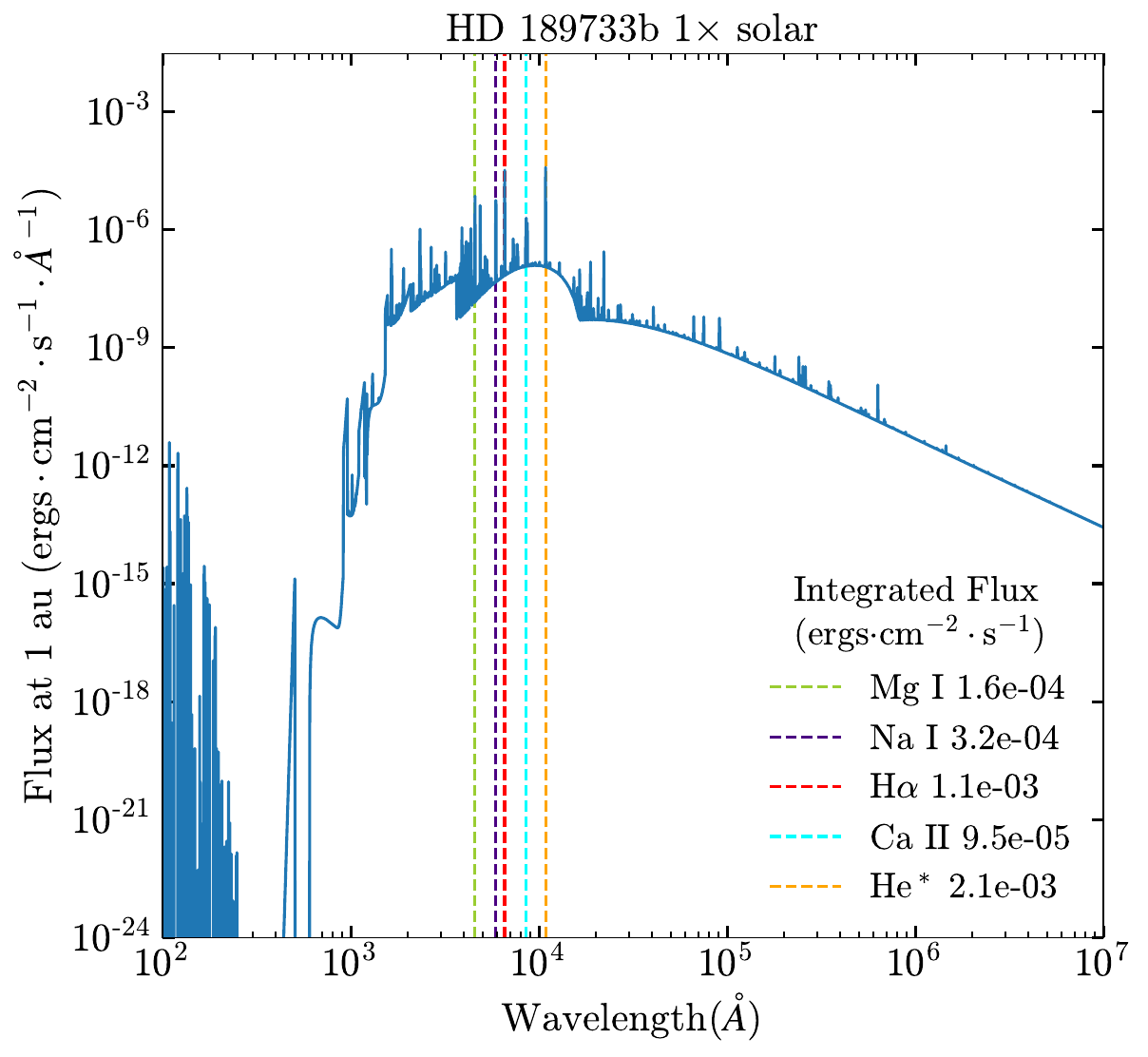}
	\caption{The emission spectrum of HD 189733b from a 1$\times$ solar metallicity model. A variety of metal lines are present, including an O I blended line at 6300 \AA.}
	\label{fig:hd189_1Z_cont}
\end{figure}

The He$^*$ triplet, 589 nm Na I doublet, and H$\alpha$ emission lines are among the brightest in all systems, as shown in Figure \ref{fig:hd189_1Z_cont}.
In runs of 1$\times$ solar metallicity or greater, we also identify bright metal lines, such as the K I doublet at 768 nm, and the Mg I forbidden line at 457 nm. The Na I doublet would be resolved by any high-resolution spectrograph, and so we calculate the maximum eclipse depths and SNRs for both lines, reporting the maximum eclipse depth and combined SNR. The Na I line is especially prominent in the HD 189733b 10$\times$ solar metallicity model, with a SNR of 2.4--higher than that of both H$\alpha$ and He$^*$.

The SNRs and maximum eclipse depths in the He$^*$ 1083 nm triplet, H$\alpha$ line, the Na I 589 nm doublet, and the [Mg I] 457 nm line are reported in Table \ref{tab:calcs}.  We also report the equivalent width of both the modeled and observed 1083 nm helium absorption signals. 

By design, the outflow from every planet we simulate has been detected in the 1083 nm helium triplet in transit.  To check the accuracy of the pyTPCI simulations, we compare the absorption signal predicted by CLOUDY with the observed absorption. The absorption signal is calculated from the number density profile of metastable state helium $n_{He*}(r)$, the temperature profile $T(r)$, and the velocity profile $v(r)$, all quantities reported by pyTPCI.

\begin{figure}
	\centering
	\includegraphics[width=0.5\textwidth]{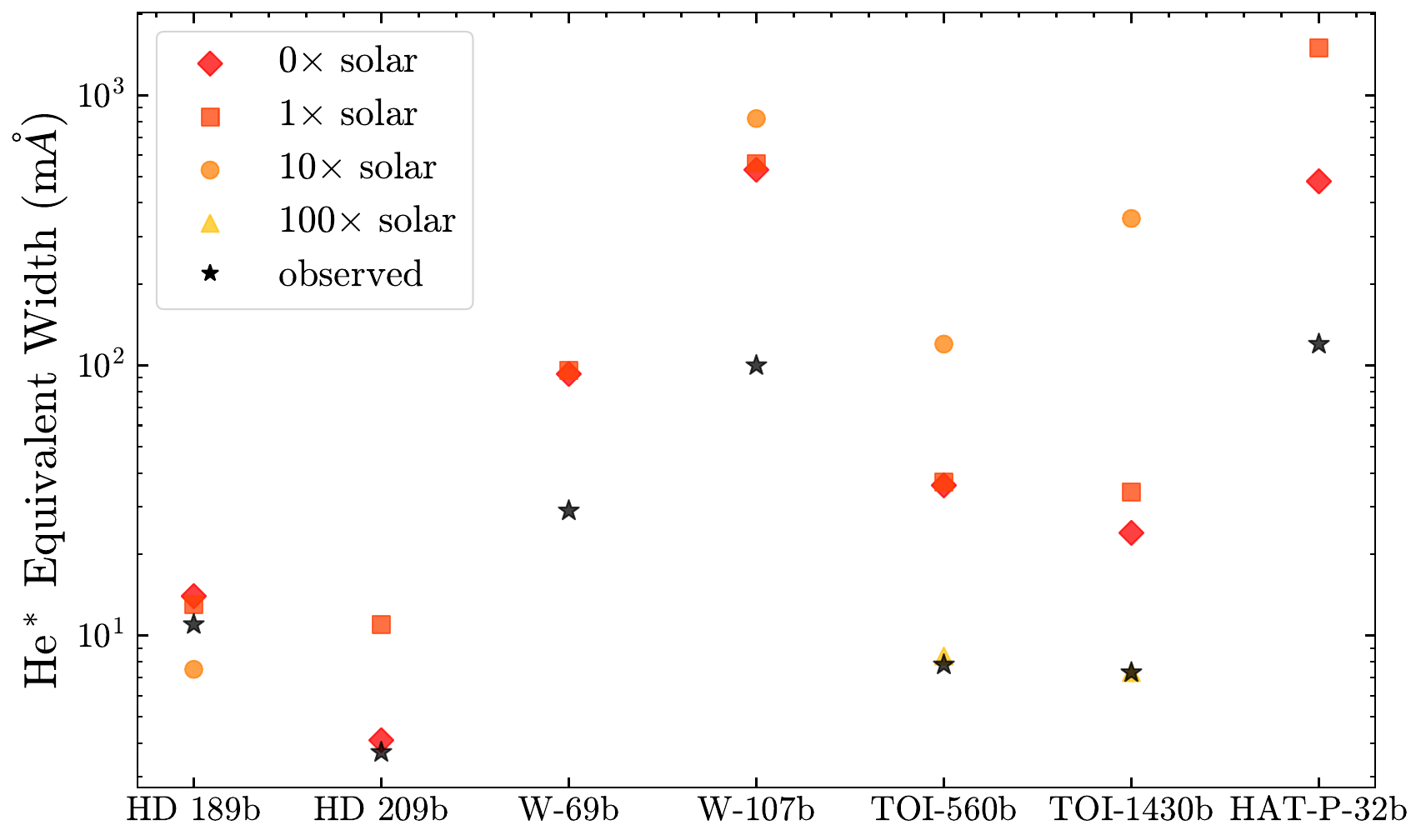}
	\caption{Predicted He$^*$ equivalent width at varying metallicities compared to the observed equivalent width. Note the good agreement of HD 189733b at 1$\times$ and 10$\times$ solar, HD 204958b at 0$\times$, and TOI-560b and TOI-1430b at 100$\times$ solar.}
	\label{fig:multiplanet_EQW}
\end{figure}

For the majority of planets, there is at least one metallicity at which the predicted He$^*$ equivalent width in absorption is very close to the observed equivalent width. For instance, the observed helium signal for HD 189733b has an equivalent width of 11 m\AA, in between our 1$\times$ solar metallicity model (13 m\AA) and our 10$\times$ solar metallicity model (7.5 m\AA).  Perhaps not coincidentally, JWST/NIRCam transmission spectra of the planet indicate a metallicity of 3--5$\times$ solar \citep{Fu2024_hd189733}. HD 209458b is fit well with zero metals, having a predicted equivalent width of 4.1 m\AA\, compared to an observed width of 3.65 m\AA. This does not agree with JWST/NIRCam transmission spectra estimating a metallicity of $4^{+5}_{-2}\times$ solar \citep{Xue24_hd209458}. The mini-Neptunes TOI-560b and TOI-1430b run at 100$\times$ solar metallicity both match their observed equivalent widths very well, with TOI-560b having a 8\% difference and TOI-1430b having a $<1$\% difference The systems of WASP-107b, WASP-69b, and HAT-P-32b are not predicted well at any metallicity, but we suspect this is a result of gas tails and extended features not being modeled properly in 1D pyTPCI, as is discussed further in Section \ref{subsec:observability}. An overview of the relationship between planet metallicity and He$^*$ equivalent width is graphed in Figure \ref{fig:multiplanet_EQW}.

\begin{deluxetable*}{ccccccccccccc}[ht]

\tabletypesize{\small}
\tablecaption{Outflow Calculations}\label{tab:calcs}
\tablehead{\colhead{Planet} & \colhead{Metallicity} & \colhead{SNR He$^*$} & \colhead{Depth} & \colhead{SNR H$\alpha$} & \colhead{Depth} & \colhead{SNR Na } & \colhead{Depth \tablenotemark{a}} & \colhead{SNR Mg} & \colhead{Depth} & \colhead{He Abs} & \colhead{W} & \colhead{{W$_{\rm obs}$}} \\ 
\colhead{} & \colhead{($\times$ solar)} & \colhead{} & \colhead{(ppm)} & \colhead{} & \colhead{(ppm)} & \colhead{} & \colhead{(ppm)} & \colhead{} & \colhead{(ppm)} & \colhead{(\%)} & \colhead{(m$\AA$)} & \colhead{(m$\AA$)} }

\startdata
HD 189733b & 0 & 0.76 & 87 & 0.49\textdagger & 69\textdagger & -- & -- & -- & -- & 2.2 & 14 & 11 \\
\textbf{HD 189733b}  & \textbf{1} & \textbf{1.3} & \textbf{170} & \textbf{0.35\textdagger} & \textbf{52\textdagger} & \textbf{0.15} & \textbf{25} & $\mathbf{6.9\e{-2}}$ \textbf{\textdagger} & \textbf{19\textdagger} & \textbf{2.0} & \textbf{13} & \textbf{11} \\
\textbf{HD 189733b} & \textbf{10} & \textbf{0.58} & \textbf{70} & \textbf{0.19\textdagger} & \textbf{86\textdagger} & \textbf{2.4} & \textbf{410} & \textbf{0.42\textdagger} & \textbf{110\textdagger} & \textbf{1.0} & \textbf{7.5} & \textbf{11} \\
\textbf{HD 209458b} & \textbf{0} & \textbf{0.23\textdagger} & \textbf{29\textdagger} & \textbf{0.25\textdagger} & \textbf{31\textdagger} & \textbf{--} & \textbf{--} & \textbf{--} & \textbf{--} & \textbf{0.65} & \textbf{4.1} & \textbf{3.7} \\
HD 209458b & 1 & 0.18\textdagger & 22\textdagger & 0.02\textdagger & 2.1\textdagger & 6.2\e{-2} & 8.5 & 0.24\textdagger & 27\textdagger & 1.5 & 11 & 3.7 \\
WASP-69b & 0 & 0.16 & 49 & 0.17\textdagger & 54\textdagger & -- & -- & -- & -- & 11 & 93 & 29 \\
WASP-69b & 1 & 0.17 & 51 & 0.15\textdagger & 43\textdagger & 0.23 & 83 & 0.19\textdagger & 74\textdagger & 12 & 96 & 29 \\
WASP-107b & 0 & 2.7 & 1400 & 0.13\textdagger & 78\textdagger & -- & -- & -- & -- & 37 & 530 & 100 \\
WASP-107b  & 1 & 2.1 & 1000 & 0.12\textdagger & 64\textdagger & 0.24 & 150 & 0.15\textdagger & 100\textdagger & 39 & 560 & 100 \\
WASP-107b & 10 & 7.1 & 3500 & 2.4\e{-3}\textdagger & 1.6\textdagger & 7.0 & 6800 & 1.1\e{-2} & 17 & 63 & 820 & 100 \\
TOI-560b & 0 & 1.8\e{-3} & 0.52 & 3.8\e{-3}\textdagger & 1.0\textdagger & -- & -- & -- & -- & 4.6 & 36 & 7.8 \\
TOI-560b  & 1 & 2.9\e{-3} & 0.84 & 3.6\e{-3}\textdagger & 0.90\textdagger & 3.3\e{-3} & 1.5 & 6.0\e{-3}\textdagger & 2.5\textdagger & 4.7 & 37 & 7.8 \\
TOI-560b  & 10 & 2.5\e{-2} & 7.1 & 4.3\e{-3}\textdagger & 1.2\textdagger & 0.18 & 47 & 6.5\e{-3}\textdagger & 2.9\textdagger & 14 & 120 & 7.8 \\
\textbf{TOI-560b}  & \textbf{100} & $\mathbf{8.0\e{-4}}$ \textbf{\textdagger} & \textbf{0.22\textdagger} & $\mathbf{8.8\e{-4}}$ \textbf{\textdagger} & \textbf{0.2\textdagger} & \textbf{0.34} & \textbf{16} & $\mathbf{1.5\e{-3}}$ \textbf{\textdagger} & \textbf{0.70\textdagger} & \textbf{1.3} & \textbf{8.4} & \textbf{7.8} \\
TOI-1430b & 0 & 3.4\e{-3} & 0.77 & 2.1\e{-3}\textdagger & 0.60\textdagger & -- & -- & -- & -- & 3.3 & 24 & 7.3 \\
TOI-1430b & 1 & 6.0\e{-3} & 1.4 & 2.4\e{-3}\textdagger & 0.60\textdagger & 9.6\e{-3} & 3.3 & 3.6\e{-3}\textdagger & 1.3\textdagger & 4.3 & 34 & 7.3 \\
TOI-1430b & 10 & 4.2\e{-2} & 9.5 & 2.1\e{-3}\textdagger & 0.50\textdagger & 0.26 & 87 & 3.2\e{-3}\textdagger & 1.1\textdagger & 37 & 350 & 7.3 \\
\textbf{TOI-1430b} & \textbf{100} & $\mathbf{4.0\e{-4}}$ \textbf{\textdagger} & \textbf{0.10\textdagger} & $\mathbf{4.4\e{-4}}$ \textbf{\textdagger} & \textbf{0.10\textdagger} & $\mathbf{3.1\e{-2}}$ & \textbf{11} & $\mathbf{7.3\e{-4}}$ \textbf{\textdagger} & \textbf{0.30\textdagger} & \textbf{1.1} & \textbf{7.3} & \textbf{7.3} \\
HAT-P-32b & 0 & 2.30 & 880 & 1.1 & 600 & -- & -- & -- & -- & 36 & 480 & 120 \\
HAT-P-32b & 1 & 11 & 3800 & 3.5 & 2000 & 0.47 & 170 & 4.5\e{-3} & 3.5 & 89 & 1500 & 120 \\
\enddata

\tablecomments{SNRs are calculated assuming a conventional eclipse observation spanning the eclipse plus an equal amount of baseline.  Under optimistic assumptions, the SNRs can be increased 2--3$\times$ by skipping the baseline and observing for several hours before or after eclipse (see Subsection \ref{subsec:boosting_snr}).}
\tablecomments{The simulations with very good fit to the observed He$^*$ equivalent width are bolded. \textdagger\, indicates that the line is optically thin.}
\tablenotetext{a}{The maximum eclipse depth between the two Na I lines are reported here.}

\end{deluxetable*}

\section{Discussion} \label{sec:discussion}

Here, we consider some implications of these pyTPCI results and what they imply about the detectability of outflow emission signals, as well as the limitations of this work, such as the simulation dimensionality, lack of molecules, lack of magnetic fields, and the uncertainty in stellar XUV flux. 

\subsection{Observability}\label{subsec:observability}
A few trends are apparent after examining Table \ref{tab:calcs}, which shows the eclipse depths and SNRs of the strongest lines. For example, higher metallicity tends to be correlated with stronger Na lines, but the outflow is sometimes so strongly suppressed at 100$\times$ solar metallicity that the line is weaker than at 10$\times$ solar metallicity, as illustrated in Figure \ref{fig:multiplanet_NaI_ED}.  This observation is also true for [Mg I] at 457 nm, but its SNRs and depths tend to be smaller, and the line is more frequently optically thin. Outflow suppression at high metallicities happens for a variety of reasons, including the onset of significant metal cooling and the decrease in easily ionizable electrons per unit mass. (For a detailed discussion, see \cite{Zhang22_toi560,Linssen24_sunbather,zhang_2024}; see also \cite{Yoshida24_molecules} about the suppressive effect of molecule cooling.)

\begin{figure}
	\centering	
    \includegraphics[width=0.5\textwidth]{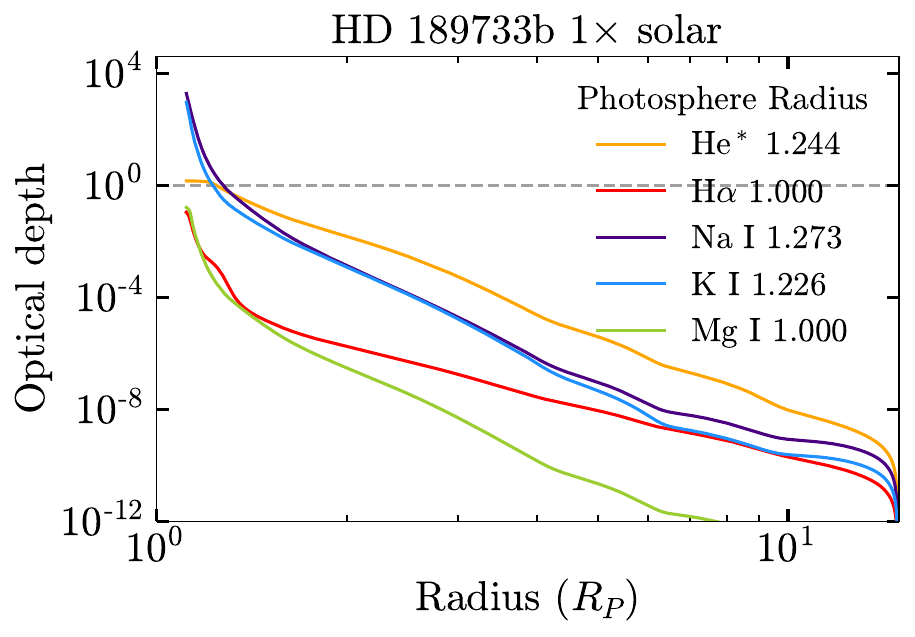}
	\caption{Optical depths of various spectral lines for HD 189733b, 1$\times$ solar metallicity. The dashed line indicates $\tau=1$.}
	\label{fig:hd189_1Z_tau}
\end{figure}

\begin{figure}
	\centering	
    \includegraphics[width=0.5\textwidth]{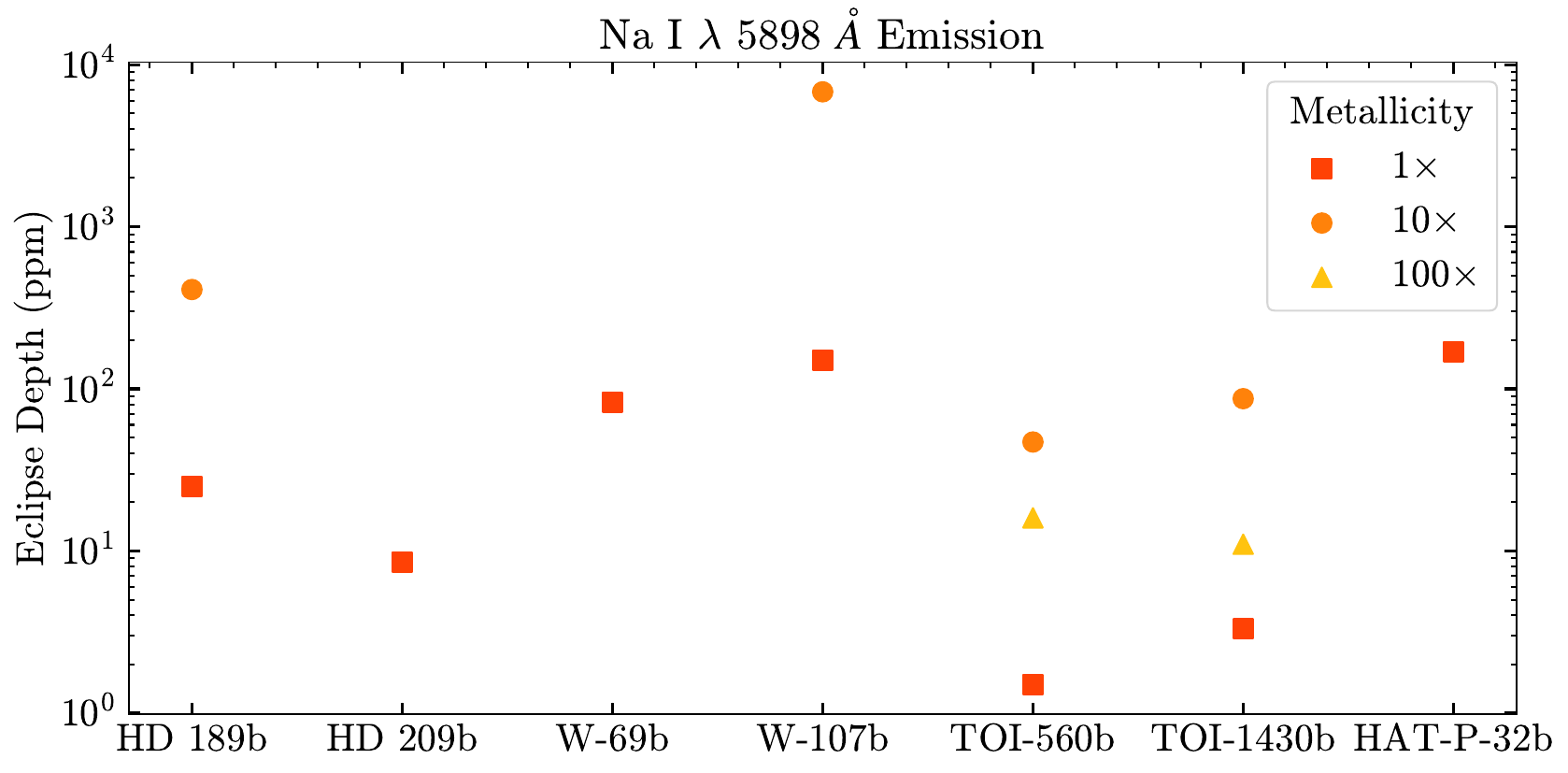}
	\caption{Eclipse depths for Na I at varying metallicities. 
 Among the metallicities we successfully tried, the eclipse depth peaks at 10$\times$ solar for every planet.  Not all systems were run at all metallicities. Refer to Table \ref{tab:calcs} for other emission lines.}
	\label{fig:multiplanet_NaI_ED}
\end{figure}

The H$\alpha$ line is nearly always optically thin (except for HAT-P-32b) and weak, with its strength not varying much with metallicity. An optically thin line indicates that some line emission could be coming from the lower hydrostatic atmosphere, not the outflow. We do not simulate the hydrostatic atmosphere, and cannot predict exactly how it will behave. However, we do expect the lower atmosphere to be much colder and have smaller velocities than the outflow, implying that lines formed in this region should be very weak and narrow. For example, if we approximate the lower-atmosphere emission from HD 189733b as a blackbody, it would have an eclipse depth less than 0.1 ppm at the H$\alpha$ wavelength, making it negligible compared to the emission from the outflow.
An example of optical depth plotted against radius for different spectral lines is presented in Figure \ref{fig:hd189_1Z_tau}, with the photosphere radius ($\tau=1$) noted for each line.

\begin{figure*}
	\centering
	\includegraphics[width=\textwidth]{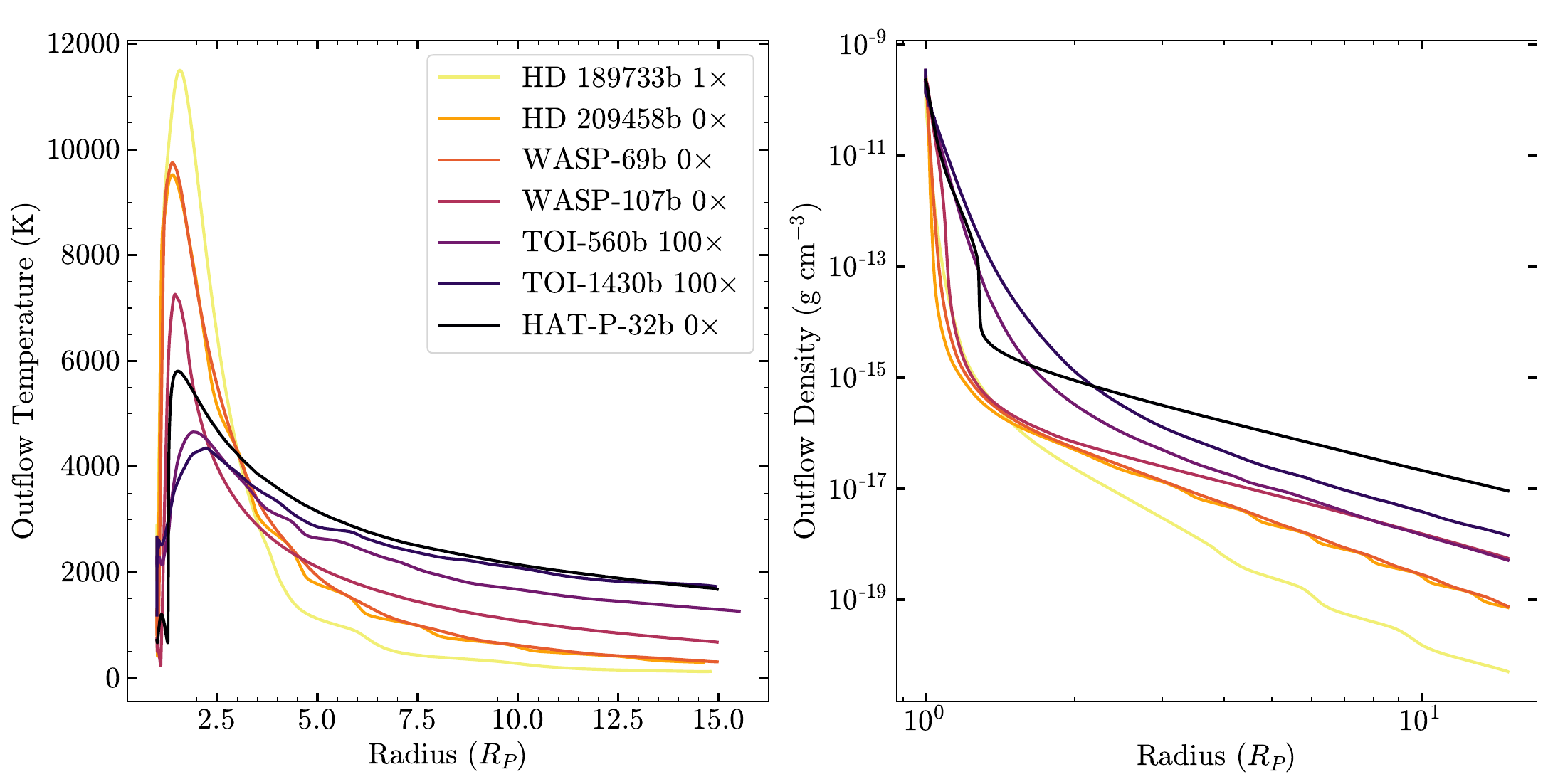}
	\caption{Outflow CLOUDY temperature and PLUTO density profiles for each system at a metallicity that best explains the observed He$^*$ equivalent width in absorption.}
	\label{fig:multiplanet_outflowprops}
\end{figure*}

The peak outflow temperature tends to decrease with metallicity and the peak outflow velocity is relatively stable across all metallicities, as tabulated in Table \ref{tab:properties} and further discussed in Section \ref{subsec:limitations}. An array of outflow temperature and density profiles are shown in Figure \ref{fig:multiplanet_outflowprops} for the systems at metallicities best matching their He$^*$ widths.

The system with the most observational potential for emission is HD 189733b, with a rather strong signal for Na I in the 10$\times$ solar metallicity run, giving a SNR of 2.4 and eclipse depth of 410 ppm.  This is the largest eclipse depth for any run that predicts a He$^*$ absorption signal close to the observed signal. HD 189733b also possesses weak but potentially observable signals for He$^*$ in its 1$\times$ and 10$\times$ solar metallicity runs. For 1$\times$ solar, we find a He$^*$ SNR of 1.3 with a maximum eclipse depth of 170 ppm, and for 10$\times$ solar, we find a SNR of 0.58 with a maximum eclipse depth of 70 ppm. H$\alpha$ is less observable for both runs, at SNRs of 0.35 and 0.19 and with eclipse depths below 100 ppm. Line emission from WASP-107b and HAT-P-32b have large SNRs ranging from 2 to 11 in He$^*$, but we believe these ought to be disregarded.

HD 189733b has the best chance of outflow emission signal detection, potentially reaching a strong SNR of 5.4 for Na I emission after observing 5 eclipses. Its SNRs for He$^*$ are somewhat weak at SNR $\sim 1$, optimistically getting a strong signal after observing about 20 eclipses---a perhaps unrealistic aim. Its H$\alpha$ SNRs are all below 1.

However, the predicted eclipse depth of $\sim 0.041\%$ for HD 189733b in Na I is very similar in magnitude to the systematic noise of $\sim 0.05\%$ on the NIRSPEC high-resolution transit spectra from helium observations of HD 189733b \citep[Fig. 2]{Zhang22_hd189733}. This indicates the Na eclipse would not be easily observable, and the He$^*$ eclipse, with a smaller predicted depth of $\sim 0.02\%$, would be even harder to observe.  Whether stacking multiple eclipses would reduce the systematic noise depends on the nature of the noise--is it independent from visit to visit, or is it highly correlated?  If the error does decrease as the square root of the number of eclipses, it may be possible to confidently detect the deeper Na I line in 5 eclipses. Comparing the high-resolution He$^*$ transit spectra of the mini-Neptunes TOI-560b and TOI-1430b \citep[Fig. 3]{Zhang23_neptunes} with the consistently tiny predicted eclipse depths and SNRs across all lines confirms that there is no chance of detecting emissions from these planets.

For systems with particularly extended exospheres or gas tails---WASP-107b \citep{Spake2018_wasp107}, WASP-69b \citep{Nortmann18_wasp69}, and HAT-P-32b \citep{Czesla22_hatp32}---the calculated SNRs appear as large as 11 for HAT-P-32b in He$^*$ or 7.1 for WASP-107b in He$^*$. How reliable are these results? We used the predicted number density profile of He$^*$ to predict the helium absorption signal and found it much higher than the observed value, indicating that these simulations should not be expected to produce realistic estimates of emission.  One possible reason is that pyTPCI, as a 1D simulation, assumes spherical symmetry. In reality, Coriolis forces would shape gas into a comet-like shape, shrinking the estimated emitting area and resultant emission signal. This is especially important for HAT-P-32b since it has a stellar radius much larger than its Coriolis turning radius, which depends on the orbital period $T_{\rm orb}$ and average outflow velocity; see Eq. \ref{eq:coriolis} and refer to Table \ref{tab:params} for values.
\begin{equation}\label{eq:coriolis}
    R_{\rm Coriolis} = \frac{v_{\rm outflow} T_{\rm orb}}{4\pi}
\end{equation}
We might expect that a Coriolis radius smaller than the stellar radius implies a violation of 1D spherical symmetry, explaining the mismatch for HAT-P-32b. However, this trend does not hold for WASP-107b and WASP-69b, which inversely have the largest Coriolis-radius-to-stellar-radius ratios out of all the systems. This could suggest different reasons for the observational-theoretical mismatch for HAT-P-32b than for WASP-107b and WASP-69b, despite the similarly large simulated equivalent widths and absorption fractions. Regardless, 3D modeling appears to be required to produce accurate results for these systems.

Na and He$^*$ emission lines are on the edge of detectability with the Keck telescopes, and so there is cause to be optimistic about the next generation of extremely large telescopes, which will possess much larger apertures coupled with state-of-the-art high-resolution spectrographs. As SNR scales linearly with aperture, even if throughput does not improve at all, the 40m diameter ELT would attain a SNR of 10 in Na and of 5.2 in He$^*$ for HD 189733b at 1$\times$ solar metallicity, a factor of four greater than the SNR achievable with Keck. This calculation makes the optimistic assumption that systematics will not be the limiting factor to the precision achievable by the ELTs.

In this work, we chose to simulate planets with detected He$^*$ absorption.  Although many of the factors that favor He$^*$ absorption detectability also favor emission detectability--such as large outflow rates, large Rp/Rs, and proximity to Earth--others (such as a K stellar type) may not.  We encourage further work on exploring the detectability of outflow emission from very different systems.  Notably, our work does not simulate any exoplanets around M dwarfs.  A natural next step is to simulate the recently discovered Jupiters orbiting M dwarfs \citep{kanodia_2024,Hotnisky24_GEMS}, which may be excellent candidates for emission searches due to their high Rp/Rs; alternatively, they may be poor candidates owing to their distance and the low luminosity of their host stars.

\subsection{Boosting SNR}
\label{subsec:boosting_snr}
The SNRs in Table \ref{tab:calcs} were calculated by assuming a standard eclipse observation: observing during the eclipse, and also during baseline on either side of eclipse.  We assumed that the total baseline duration is equal to the eclipse duration.  We can gain a factor of nearly $\sqrt{2}$ in SNR by observing a very long baseline and assuming (very optimistically) that there is no significant stellar, instrumental, or telluric variability during this interval.

Another way to boost the SNR is to adopt a different observation strategy: observe the planet either before or after the eclipse for much longer than the eclipse duration.  The planet signal would then be disentangled from the stellar spectrum by relying on the radial acceleration of the planet.  The mean spectrum would be subtracted from each individual spectrum, and principal component analysis could be used to subtract off telluric and stellar variability.  The cleaned residual spectra could then be shifted into the planet frame and stacked.  This strategy is very similar to high-resolution cross-correlation spectroscopy, except that cross correlation is not necessary.  It has the potential to boost SNR significantly.  For example, HD 189733b has an eclipse duration of 1.8 h, but HD 189733 is observable for 8.5 hours from Keck at airmasses below 2.  This shift-and-stack approach can potentially boost the SNR by a factor of $\sqrt{8.5/1.8}\sqrt{2} = 3$ (the factor of $\sqrt{2}$ is because one no longer needs to compare to a baseline).  Under this scenario, the Na emission from HD 189733b would be detectable at 7.2$\sigma$ (assuming $10\times$ solar metallicity).

In practice, there are weaknesses with the shift-and-stack approach that diminish the SNR gain.  8.5 h is likely overly optimistic because it is hard to find nights where HD 189733b is near eclipse and also at low airmass for 8.5 hours.  The airmass, star, instrument, and Earth's atmosphere all change more on 8-hour timescales than 3.6-hour timescales.  PCA can subtract off most of the variability, but it also subtracts some percentage of the signal.  It is also not clear that the radial velocity of the outflow emission lines follows the radial velocity of the planet.  In absorption, the 1083 nm helium line often diverges significantly from the planet's radial velocity track due to outflow morphology, such as comet-like tails (WASP-107b is a striking example; see \citealt{kirk_2020}).  All in all, it is plausible that one could see a SNR gain from this alternative observation strategy, but it is riskier. One other possible benefit from this strategy is that if--contrary to our predictions--the emission signal is very strong, measuring the emission spectrum over time could provide kinematic information about the outflow structure (as it does for 1083 nm helium absorption measurements).

\subsection{Limitations}\label{subsec:limitations}
Mainly intended as a an order-of-magnitude investigation into observability, there are substantial limitations of this work. On one hand, the fact that pyTPCI is 1D fundamentally limits some of the phenomena that can be modeled, such as the asymmetric line profiles of systems with gas tails like WASP-69b, which require a 2D simulation to capture day-night advection and a 2.5D simulation to include Coriolis forces \citep{Wang21_wasp69}. On the other hand, pyTPCI's dimensionality allows it to include far more detailed atomic physics and radiative transfer, meaning it has complementary strengths and weaknesses to 3D models.

CLOUDY possesses the ability to run spectral synthesis with molecules enabled, including species like H$_{2}$ and CO, but all of our pyTPCI simulations were run with molecules turned off due to issues with stability and convergence. It is possible that some molecules are entrained in the outflow at the pressures probed by the simulation domain, and it is worth noting that the presence of molecules could drive substantial cooling of the outflow, suppressing the mass loss rate \citep{Yoshida24_molecules}.
Fortunately, expanding this study with molecules enabled should be theoretically possible. 

Additionally, CLOUDY does not take into account mass fractionation---it assumes a hydrodynamic outflow in which all species remain well mixed, but in reality, lighter elements are preferentially lost. This likely results in overestimating the emission from metal lines like Na.  However, mass fractionation is only significant for weak outflows, which would be unfavorable emission targets anyhow. We briefly demonstrate that mass fractionation is usually unimportant by comparing the diffusion-limited escape rate and the estimated escape rate. Following \cite{Zhang24_toi836}, we find that the diffusion-limited mass loss rate can be approximated as the following, for a planet mass $M_p$, outflow temperature $T$, and mean molecular weight $\mu$:
\begin{equation}\label{eq:DL}
    \Phi_{DL} = (1.3 \times 10^{8} \rm g \cdot \rm s^{-1}) \Big(\frac{M_p}{M_\earth}\Big) \Big(\frac{T}{10^3 \rm K}\Big)^{-0.25} \mu
\end{equation}

As an example of a hot Jupiter, for HD 209458b at solar metallicity, we find a diffusion-limited mass rate of $2.0 \times 10^{10} \rm g\, \rm s^{-1}$, compared to the simulated mass loss rate $ 2.7 \times 10^{11} \rm g\, \rm s^{-1}$ and an estimated value spanning $0.42-1.00\e{11}\rm g\, \rm s^{-1}$ \citep{Lampon21_escobs}. For most simulations, the estimated mass loss rate is much larger than the diffusion-limited mass loss rate. However, the estimated mass loss rate for TOI-1430b at 100$\times$ solar is only 3$\times$ greater than the diffusion-limited rate, and for TOI-560b at 100$\times$ solar is only 1.5$\times$ greater than the diffusion-limited rate. For HD 189733b at 10$\times$ solar metallicity, we find a diffusion-limited mass loss rate of $3.8 \times 10^{10} \rm g\, \rm s^{-1}$, about 200$\times$ greater than the simulated mass loss rate. This holds for all HD 189733b runs, though the highest metallicity run most badly violates the hydrodynamic assumption. These diffusion-limited calculations assume that the outflow is neutral and do not consider the high ionization rate at higher radii which would occur in a true outflow. This would substantially lower the diffusion-limited mass loss rate because ion-ion interactions are far stronger than neutral-neutral interactions. HD 189733b in particular becomes highly ionized at especially low radii. For 1$\times$ solar metallicity, 10\% of hydrogen is ionized at 1.26 planetary radii, and 50\% ionized at 1.28 planetary radii. More work remains in order to conclude whether diffusive separation is indeed significant for HD 189733b. The simulated mass loss rate, diffusion-limited mass loss rate, estimated mass loss rate from the literature, and peak outflow temperature and velocity are reported in Table \ref{tab:properties}.

Mass loss rate estimates for HD 189733b vary widely, but neither the HD 189733b 1$\times$ solar model at $5.7\e{9} \rm g\, \rm s^{-1}$ nor the 10$\times$ solar model at $1.9\e{8} \rm g\, \rm s^{-1}$ agree well with $1.1\e{11} \rm g\, \rm s^{-1}$ from \citet{Lampon21_hd189733}. However, the 1$\times$ solar model agrees well with a different estimate of $5\e{9} \rm g\, \rm s^{-1}$ from \citet{Zhang22_hd189733}. The mass loss rate for HD 209458b 0$\times$ falls within the range established in \citet{Lampon21_escobs}. For TOI-560b 100$\times$ and TOI-1430b 100$\times$, our mass loss rates are smaller than the estimates of \citet{Zhang23_neptunes} by an order of magnitude, likely because of the much higher metallicities. Though we do not have models of WASP-69b, WASP-107b, and HAT-P-32b with good agreement to the observed He$^*$ equivalent width, the simulated mass loss rates generally fall near other literature estimates \citep{Wang21_wasp69, Spake2018_wasp107, Lampon23_he}.

\begin{deluxetable*}{ccccccc}[ht]

\tablecaption{Outflow Properties}\label{tab:properties}

\tablehead{\colhead{System} & \colhead{Metallicity} & \colhead{$\dot{m}_{\rm sim}$} & \colhead{$\dot{m}_{DL}$} & \colhead{$\dot{m}_{\rm lit}$} & \colhead{T$_{\rm outflow}$} & \colhead{v$_{\rm outflow}$} \\ 
\colhead{} & \colhead{($\times$ solar)} & \colhead{(g/s)} & \colhead{(g/s)} & \colhead{(g/s)} & \colhead{(K)} & \colhead{(km/s)} } 

\startdata
HD 189733b & 0 & 1.1\e{10} & 3.2\e{10} & 1.1\e{11} & 12,400 & 70 \\
\textbf{HD 189733b}  & \textbf{1} & $\mathbf{5.7\e{9}}$ & $\mathbf{3.2\e{10}}$ & $\mathbf{1.1\e{11}}$ & \textbf{11,500} & \textbf{70} \\
\textbf{HD 189733b} & \textbf{10} & $\mathbf{1.9\e{8}}$ & $\mathbf{3.8\e{10}}$ & $\mathbf{1.1\e{11}}$ & \textbf{9,700} & \textbf{71} \\
\textbf{HD 209458b} & \textbf{0} & $\mathbf{1.1\e{11}}$ & $\mathbf{2.0\e{10}}$ & $\mathbf{0.42-1.00\e{11}}$ & \textbf{9,500} & \textbf{55} \\
HD 209458b & 1 & 2.7\e{11} & 2.0\e{10} & 0.42-1.00\e{11} & 8,900 & 55 \\
WASP-69b & 0 & 7.5\e{10} & 8.3\e{9} & 1\e{11} & 9,700 & 55 \\
WASP-69b & 1 & 7.2\e{10} & 8.5\e{9} & 1\e{11} & 9,000 & 54 \\
WASP-107b & 0 & 2.1\e{11} & 3.7\e{9} & 1\e{10}--3\e{11} & 7,200 & 30 \\
WASP-107b  & 1 & 2.3\e{11} & 3.7\e{9} & 1\e{10}--3\e{11} & 7,200 & 30 \\
WASP-107b & 10 & 9.2\e{11} & 4.2\e{9} & 1\e{10}--3\e{11} & 6,900 & 31 \\
TOI-560b & 0 & 5.4\e{9} & 1.2\e{9} & 2.3\e{10} & 4,000 & 11 \\
TOI-560b  & 1 & 5.6\e{9} & 1.3\e{9} & 2.3\e{10} & 3,900 & 11 \\
TOI-560b  & 10 & 1.9\e{10} & 1.5\e{9} & 2.3\e{10} & 3,400 & 12 \\
\textbf{TOI-560b}  & \textbf{100} & $\mathbf{3.7\e{9}}$ & $\mathbf{2.5\e{9}}$ & $\mathbf{2.3\e{10}}$ & \textbf{4,700} & \textbf{8.1} \\
TOI-1430b & 0 & 3.9\e{9} & 8.4\e{8} & 2.6\e{10} & 3,100 & 8.4 \\
TOI-1430b & 1 & 5.6\e{9} & 8.5\e{8} & 2.6\e{10} & 3,000 & 9.3 \\
TOI-1430b & 10 & 4.4\e{10} & 9.6\e{8} & 2.6\e{10} & 3,000 & 11 \\
\textbf{TOI-1430b} & \textbf{100} & $\mathbf{5.4\e{9}}$ & $\mathbf{1.7\e{9}}$ & $\mathbf{2.6\e{10}}$ & \textbf{4,400} & \textbf{7.9} \\
HAT-P-32b & 0 & 1.1\e{13} & 2.2\e{10} & 1.30\e{13} & 5,800 & 22 \\
HAT-P-32b & 1 & 5.4\e{13} & 2.2\e{10} & 1.30\e{13} & 6,500 & 25 \\
\enddata

\tablecomments{The simulations with very good fit to the observed He$^*$ equivalent width are bolded. The listed temperatures and velocities are the peak values.}
\tablerefs{Estimated mass loss rates from the literature. HD 189733b: \citet{Lampon21_hd189733}; HD 209458b: \citet{Lampon21_escobs}; WASP-69b: \citet{Wang21_wasp69}; WASP-107b: \citet{Spake2018_wasp107}; TOI-560b: \citet{Zhang23_neptunes}; TOI-1430b: \citet{Zhang23_neptunes}; HAT-P-32b: \citet{Lampon23_he}.}

\end{deluxetable*}

Another limitation of our simulations is that we do not systematically consider the effects of magnetic fields, which \citet{Adams11_HotJupiters} suggests should significantly impact the outflow geometry of hot Jupiters. This is validated by 3D radiative magnetohydrodynamic simulations (e.g. \citet{Arakcheev17_wasp12, Owen19_valleymag, Carolan21_obsmagnetic, Schreyer24_mag}). In general, the problem of simulating outflow magnetic fields remains to be explored at depth, but most studies agree that the presence of a planetary magnetic field would reduce mass loss rates \citep{Brain24_exoplanetmagneticfields}.

Additionally, X-ray and EUV spectra of a system's host star are necessary for obtaining precise number densities and mass loss rates, but in most cases these fluxes remain uncertain within a factor of at least a few, as discussed in \citet{France22_EUV}. Some systems possess XMM-Newton X-ray measurements, but no telescope capable of observing stellar EUV emission is currently in operation. The impact of varying XUV stellar spectra is briefly examined in Appendix \ref{sec:apptab}, but usually higher XUV flux results in a higher eclipse depth and SNR.

Finally, this work also leaves some of the high-metallicity parameter space unexplored due to stability concerns with pyTPCI. For example, based on our current equivalent width results we do not think it particularly likely that a system like WASP-69b would be best fit by 10$\times$ solar metallicity, but it would still be appropriate to test that possibility regardless. 

\section{Conclusion} \label{sec:conclusion}
In this work, we examine the detectability of emissions from exoplanet outflows using the new radiative-hydrodynamics code pyTPCI. Using it,  we simulate seven exoplanets: two hot Jupiters (HD 189733b and HD 209458b), two Saturnian exoplanets (WASP-69b and WASP-107b), two mini-Neptunes (TOI-560b and TOI-1430b) and one ultra-hot Jupiter (HAT-P-32b) at varying metallicities ranging from 0$\times$ to 100$\times$ solar. 

We introduce various features to stabilize TPCI \citep{Salz15_TPCI} and increase its ease of use, and additionally take measures to increase accuracy in a 1D simulation, such as setting the illumination angle to 66$^\circ$. We have released pyTPCI on GitHub (\url{https://github.com/ideasrule/pyTPCI}) where it is available for everyone to use, and will be further maintained. 

We obtain continuum emission spectra from pyTPCI and calculate the eclipse depths and signal-to-noise ratios for various spectral lines. We perform these detectability calculations for a 10m class telescope with a 10\% throughput high-resolution spectrograph, similar to the specifications of KPF or NIRSPEC on Keck. We determine simulation accuracy by comparing the metastable helium absorption equivalent width predicted by our simulations with the observed equivalent widths. We get simulations that predict helium equivalent widths very close to the observed values for HD 189733b, HD 209458b, TOI-560b, and TOI-1430b.  We believe the emission predictions for these planets are the most trustworthy.

We find that the most detectable planet simulated is HD 189733b, with the most detectable lines being the 589 nm Na doublet with a eclipse depth of 410 ppm and SNR of 2.4, and the metastable helium triplet with an eclipse depth of 170 ppm and a SNR of 1.3. These signatures would only be marginally detectable with Keck with multiple eclipse observations, but they would be much more detectable using the next generation of extremely large telescopes.

Our simulations are limited by their lack of magnetic fields, stellar wind, day-night inhomogeneity, and molecules. They do not include mass fractionation, though we do not expect diffusive separation to be significant in most systems besides perhaps HD 189733b. pyTPCI---like all mass loss simulations---requires X-ray and EUV spectra of the host star, though these are frequently uncertain to factors of at least a few. Some systems like WASP-69b, WASP-107b, and HAT-P-32b likely require 3D simulations to fully account for the asymmetric and extended outflow features.

In order to put these results to the test, we encourage further modeling as well as observations of exoplanet outflow emissions. Observations of these outflow emissions would enable further development of models, improving our understanding of the mass loss mechanisms that likely help carve the radius valley.

\begin{acknowledgments}
This work was funded by XMM-Newton grant 80NSSC24K0333. We thank Jorge Sanz-Forcada for providing XUV spectra for our selection of planets before the publication of \citet{SanzForcada25_He}. MZ thanks the Heising-Simons Foundation for funding his 51 Pegasi b fellowship. RR thanks Fausto Cattaneo and Leslie Rogers for their feedback on prior iterations of this project.
\end{acknowledgments}

%

\vspace{5mm}


\software{pyTPCI \citep{Zhang_Rosener_2024},
          TPCI \citep{Salz15_TPCI},
          Cloudy 23 \citep{Chatzikos23_CL23}, 
          PLUTO 4.4 \citep{Mignone07_PLUTO},
          PHOENIX \citep{Husser13_PHOENIX}
          }



\appendix

\section{Outflow Simulations Using Alternative XUV Spectra}\label{sec:apptab}
In earlier stages, pyTPCI was run using stellar spectra constructed by the method described in \citet{Salz16_escaping}, as opposed to the XUV spectra from \citet{SanzForcada25_He}.
Table \ref{tab:appcalcs} reports results from selected earlier runs of pyTPCI. Some of them were run at an illumination angle of 0$^\circ$ instead of 66$^\circ$ suggested by \citet{johnstone_2018}, to explore the effects of this simplification.

In Figure \ref{fig:0Z_salz_comp}, we compare pyTPCI runs of HD 189733b and HD 209458b with the TPCI results that \citet{Salz16_escaping} obtained.  We performed both pyTPCI runs at 0$\times$ solar metallicity at 0$^\circ$ irradiation angle, just as in \cite{Salz16_escaping}.  This shows excellent agreement between the two systems, as we should expect for a similar simulation setup and identical stellar spectra. 

\begin{figure*}[h]
    \centering
    \includegraphics[width=0.6\linewidth]{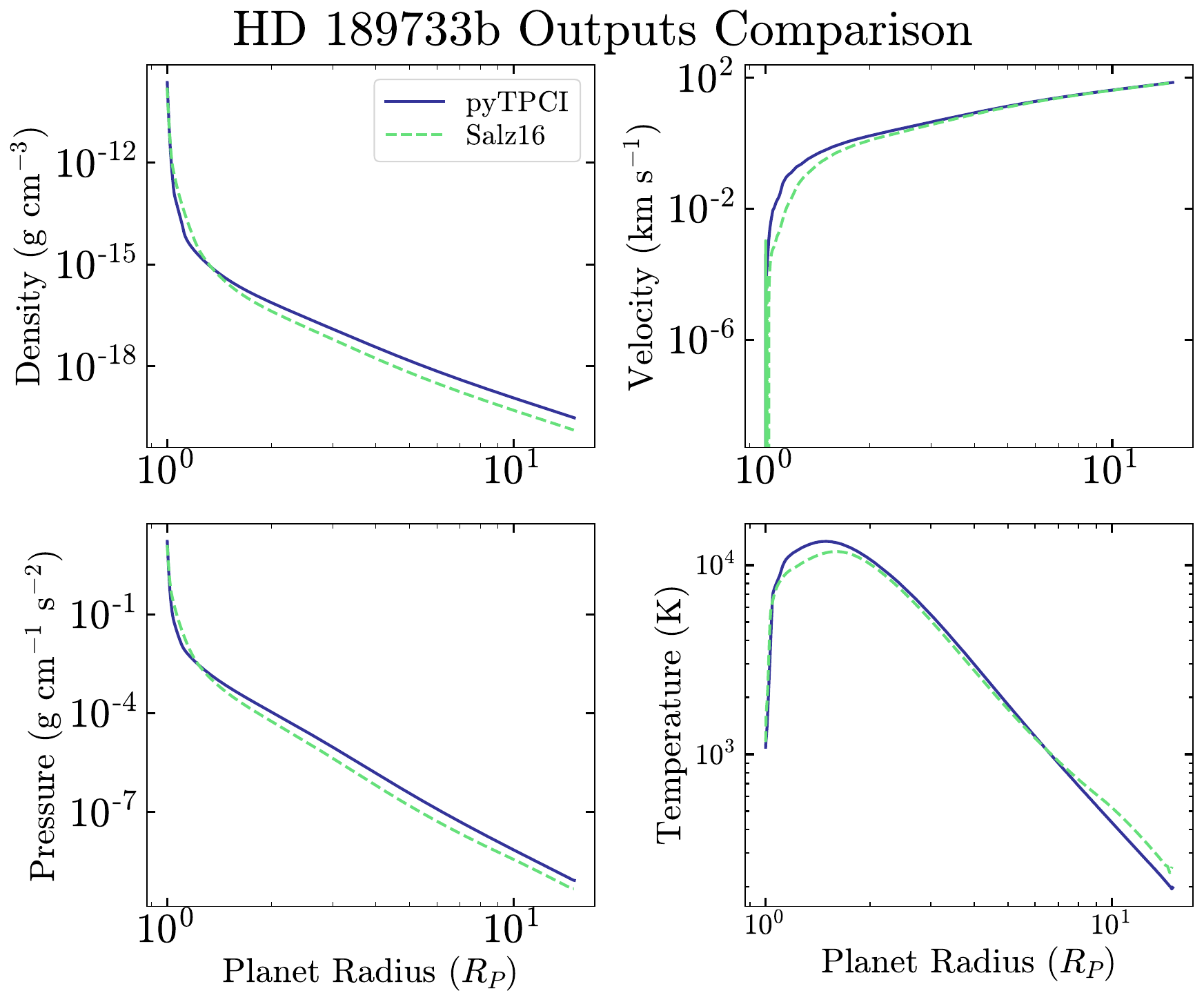}
    \caption{Data outputs for HD 189733b pyTPCI simulations compared to those of \citet{Salz16_escaping}, run at 0$\times$ solar metallicity and 0$^\circ$ illumination angle.}
    \label{fig:0Z_salz_comp}
\end{figure*}
\begin{figure*}[h]
    \centering
    \includegraphics[width=0.6\linewidth]{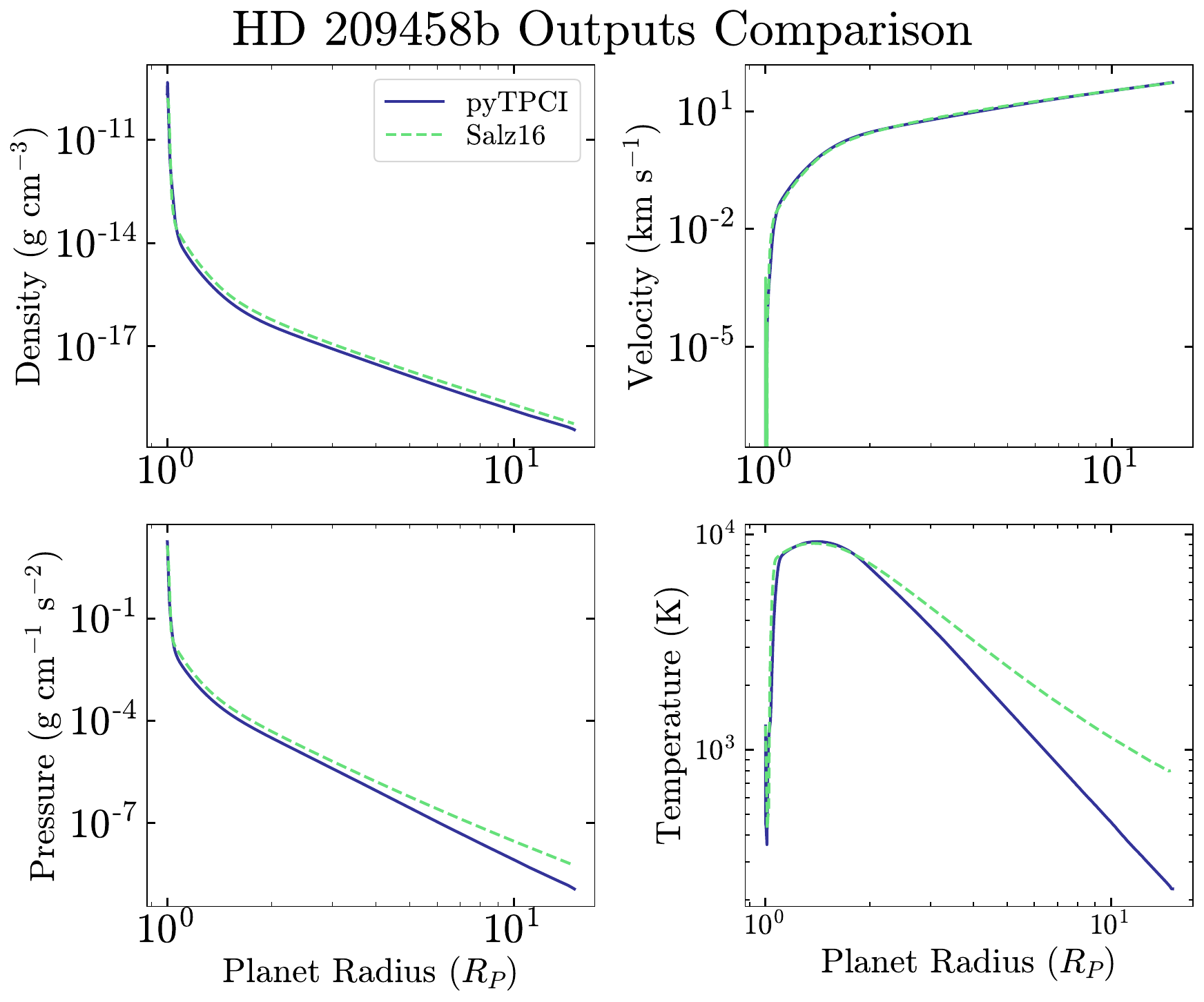}
    \caption{Data outputs for HD 209458b pyTPCI simulations compared to those of \citet{Salz16_escaping}, run at 0$\times$ solar metallicity and 0$^\circ$ illumination angle.}
    \label{fig:0Z_salz_comp}
\end{figure*}

As expected, the eclipse depths and SNRs for 0$^\circ$ illumination angle runs are larger. For example, the HD 209458b 0$\times$ solar metallicity run has an eclipse depth of 400 ppm and SNR of 0.75 in He$^*$ at the substellar point, compared with an eclipse depth of 24 ppm and SNR of 0.19 at 66$^\circ$. 

For the HD 189733b stellar spectrum, the XUV flux (1--504 \AA) at 1 AU of the previous spectrum was 13 ergs$\cdot$s$^{-1} \cdot$ cm$^{-2}$, compared with a flux of 18 ergs$\cdot$s$^{-1} \cdot$ cm$^{-2}$ for the \citet{SanzForcada25_He} spectrum. These flux differences are significant but not atypical for high-energy stellar spectra. Notably, most of the Sanz-Forcada spectra (except in the case of HD 189733b) have lower XUV fluxes than the corresponding Salz spectra by at least 40\%.
Generally, higher XUV flux results in higher eclipse depths and SNRs. For example, for WASP-107b 0$\times$ solar, the model using the the Salz stellar spectrum with 60\% greater flux has a He$^*$ eclipse depth and SNR 50\% greater than the model with the Sanz-Forcada spectrum: 1400ppm and 2.7 for the former compared to 4.0 and 2100ppm for the latter. However, this trend is broken by some of the TOI-560b and TOI-1430b runs.

The mechanisms producing the impossibly large He$^*$ absorption fractions and equivalent widths for WASP-69b, WASP-107b, and HAT-P-32b described in Section \ref{subsec:observability} are also at work here, as evidenced by the eclipse depth of 13000 ppm and SNR of 27 for HAT-P-32b at 1$\times$ solar, as the most drastic example.

\begin{deluxetable}{cccccccccc}

\tablecaption{Outflow Simulations Using Alternative XUV Spectra}\label{tab:appcalcs}

\tablehead{\colhead{Planet} & \colhead{Metallicity} & \colhead{SNR He$^*$} & \colhead{Depth} & \colhead{SNR H$\alpha$} & \colhead{Depth} & \colhead{He Abs} & \colhead{W} & \colhead{W$_{\rm obs}$} & \colhead{T$_{\rm peak}$} \\ 
\colhead{} & \colhead{($\times$ solar)} & \colhead{} & \colhead{(ppm)} & \colhead{} & \colhead{(ppm)} & \colhead{(\%)} & \colhead{(m\AA)} & \colhead{(m\AA)} & \colhead{(K)} } 

\startdata
HD 189733b & 0 & 1.7 & 190 & 1.6 & 250 & 4.4 & 29 & 11 & 13,400 \\
HD 189733b & 1 & 1.7 & 190 & 1.1 & 180 & 2.7 & 19 & 11 & 11,800 \\
HD 189733b & 10 & 1.8 & 210 & 0.32 & 44 & 1.3 & 10 & 11 & 10,300 \\
HD 189733b \textdaggerdbl & 0 & 1.9 & 610 & 1.2 & 180 & 3.2 & 22 & 11 & 11,600 \\
HD 209458b & 0 & 0.19 & 24 & 0.41 & 57 & 0.66 & 3.9 & 3.7 & 9,300 \\
HD 209458b & 1 & 0.18 & 23 & 0.34 & 44 & 1.5 & 9.3 & 3.7 & 8,400 \\
HD 209458b \textdaggerdbl & 0 & 0.60 & 76 & 0.42 & 58 & 1.5 & 9.1 & 3.7 & 8,800 \\
WASP-69b & 0 & 0.15 & 41 & 0.19 & 71 & 9.0 & 76 & 29 & 9,600 \\
WASP-69b & 1 & 0.35 & 170 & 0.18 & 76 & 20 & 220 & 29 & 7,900 \\
WASP-69b \textdaggerdbl & 0 & 0.35 & 96 & 0.30 & 110 & 16 & 160 & 29 & 9,300 \\
WASP-107b & 0 & 4.0 & 2100 & 7.7\e{-2} & 62 & 34 & 510 & 100 & 8,600 \\
WASP-107b & 1 & 3.8 & 2000 & 5.4\e{-2} & 42 & 37 & 540 & 100 & 8,500 \\
WASP-107b & 10 & 3.5 & 1800 & 3.3\e{-2} & 22 & 49 & 610 & 100 & 7,800 \\
WASP-107b \textdaggerdbl & 0 & 1.5 & 740 & 1.9\e{-2} & 25 & 80 & 610 & 100 & 6,100 \\
TOI-560b & 0 & 0.15 & 42 & 4.2\e{-3} & 1.3 & 9.6 & 99 & 7.8 & 4,900 \\
TOI-560b & 100 & 2.3\e{-3} & 0.61 & 3.3\e{-4} & 0.10 & 1.2 & 8.5 & 7.8 & 4,800 \\
TOI-560b \textdaggerdbl & 0 & 0.18 & 51 & 1.7\e{-3} & 0.54 & 45 & 280 & 7.8 & 4,100 \\
TOI-1430b & 0 & 5.8\e{-2} & 14 & 2.3\e{-3} & 0.72 & 5.8 & 66 & 7.3 & 5,100 \\
TOI-1430b & 100 & 8.2\e{-4} & 0.19 & 1.3\e{-4} & 0.034 & 0.96 & 7.4 & 7.3 & 4,600 \\
TOI-1430b \textdaggerdbl & 0 & 2.4\e{-3} & 0.50 & 1.2\e{-3} & 0.35 & 17 & 93 & 7.3 & 7,900 \\
HAT-P-32b & 0 & 4.7 & 2500 & 2.4 & 1400 & 52 & 470 & 120 & 4,400 \\
HAT-P-32b & 1 & 27 & 13000 & 0.84 & 450 & 81 & 630 & 120 & 6,100 \\
\enddata

\tablecomments{Simulations with \textdaggerdbl\, indicate 0$^\circ$ illumination angle. These data are not the final results.}

\end{deluxetable}

\bibliography{mainbib}{}

\begin{thebibliography}{}
\expandafter\ifx\csname natexlab\endcsname\relax\def\natexlab#1{#1}\fi
\providecommand{\url}[1]{\href{#1}{#1}}
\providecommand{\dodoi}[1]{doi:~\href{http://doi.org/#1}{\nolinkurl{#1}}}
\providecommand{\doeprint}[1]{\href{http://ascl.net/#1}{\nolinkurl{http://ascl.net/#1}}}
\providecommand{\doarXiv}[1]{\href{https://arxiv.org/abs/#1}{\nolinkurl{https://arxiv.org/abs/#1}}}

\bibitem[{Adams(2011)}]{Adams11_HotJupiters}
Adams, F.~C. 2011, The Astrophysical Journal, 730, 27, \dodoi{10.1088/0004-637x/730/1/27}

\bibitem[{{Addison} {et~al.}(2019){Addison}, {Wright}, {Wittenmyer}, {Horner}, {Mengel}, {Johns}, {Marti}, {Nicholson}, {Soutter}, {Bowler}, {Crossfield}, {Kane}, {Kielkopf}, {Plavchan}, {Tinney}, {Zhang}, {Clark}, {Clerte}, {Eastman}, {Swift}, {Bottom}, {Muirhead}, {McCrady}, {Herzig}, {Hogstrom}, {Wilson}, {Sliski}, {Johnson}, {Wright}, {Johnson}, {Blake}, {Riddle}, {Lin}, {Cornachione}, {Bedding}, {Stello}, {Huber}, {Marsden}, \& {Carter}}]{Addison19_hd189733}
{Addison}, B., {Wright}, D.~J., {Wittenmyer}, R.~A., {et~al.} 2019, \pasp, 131, 115003, \dodoi{10.1088/1538-3873/ab03aa}

\bibitem[{{Arakcheev} {et~al.}(2017){Arakcheev}, {Zhilkin}, {Kaigorodov}, {Bisikalo}, \& {Kosovichev}}]{Arakcheev17_wasp12}
{Arakcheev}, A.~S., {Zhilkin}, A.~G., {Kaigorodov}, P.~V., {Bisikalo}, D.~V., \& {Kosovichev}, A.~G. 2017, Astronomy Reports, 61, 932, \dodoi{10.1134/S1063772917110014}

\bibitem[{{Barrag{\'a}n} {et~al.}(2022){Barrag{\'a}n}, {Armstrong}, {Gandolfi}, {Carleo}, {Vidotto}, {Villarreal D'Angelo}, {Oklop{\v{c}}i{\'c}}, {Isaacson}, {Oddo}, {Collins}, {Fridlund}, {Sousa}, {Persson}, {Hellier}, {Howell}, {Howard}, {Redfield}, {Eisner}, {Georgieva}, {Dragomir}, {Bayliss}, {Nielsen}, {Klein}, {Aigrain}, {Zhang}, {Teske}, {Twicken}, {Jenkins}, {Esposito}, {Van Eylen}, {Rodler}, {Adibekyan}, {Alarcon}, {Anderson}, {Akana Murphy}, {Barrado}, {Barros}, {Benneke}, {Bouchy}, {Bryant}, {Butler}, {Burt}, {Cabrera}, {Casewell}, {Chaturvedi}, {Cloutier}, {Cochran}, {Crane}, {Crossfield}, {Crouzet}, {Collins}, {Dai}, {Deeg}, {Deline}, {Demangeon}, {Dumusque}, {Figueira}, {Furlan}, {Gnilka}, {Goad}, {Goffo}, {Guti{\'e}rrez-Canales}, {Hadjigeorghiou}, {Hartman}, {Hatzes}, {Harris}, {Henderson}, {Hirano}, {Hojjatpanah}, {Hoyer}, {Kab{\'a}th}, {Korth}, {Lillo-Box}, {Luque}, {Marmier}, {Mo{\v{c}}nik}, {Muresan}, {Murgas}, {Nagel}, {Osborne}, {Osborn}, {Osborn}, {Palle}, {Raimbault}, {Ricker},
  {Rubenzahl}, {Stockdale}, {Santos}, {Scott}, {Schwarz}, {Shectman}, {Raimbault}, {Seager}, {S{\'e}gransan}, {Serrano}, {Skarka}, {Smith}, {{\v{S}}ubjak}, {Tan}, {Udry}, {Watson}, {Wheatley}, {West}, {Winn}, {Wang}, {Wolfgang}, \& {Ziegler}}]{Barragan21_toi560}
{Barrag{\'a}n}, O., {Armstrong}, D.~J., {Gandolfi}, D., {et~al.} 2022, \mnras, 514, 1606, \dodoi{10.1093/mnras/stac638}

\bibitem[{{Bonomo} {et~al.}(2017){Bonomo}, {Desidera}, {Benatti}, {Borsa}, {Crespi}, {Damasso}, {Lanza}, {Sozzetti}, {Lodato}, {Marzari}, {Boccato}, {Claudi}, {Cosentino}, {Covino}, {Gratton}, {Maggio}, {Micela}, {Molinari}, {Pagano}, {Piotto}, {Poretti}, {Smareglia}, {Affer}, {Biazzo}, {Bignamini}, {Esposito}, {Giacobbe}, {H{\'e}brard}, {Malavolta}, {Maldonado}, {Mancini}, {Martinez Fiorenzano}, {Masiero}, {Nascimbeni}, {Pedani}, {Rainer}, \& {Scandariato}}]{Bonomo17_hd209458}
{Bonomo}, A.~S., {Desidera}, S., {Benatti}, S., {et~al.} 2017, \aap, 602, A107, \dodoi{10.1051/0004-6361/201629882}

\bibitem[{Brain {et~al.}(2024)Brain, Kao, \& O'Rourke}]{Brain24_exoplanetmagneticfields}
Brain, D.~A., Kao, M.~M., \& O'Rourke, J.~G. 2024, Exoplanet Magnetic Fields.
\newblock \doarXiv{2404.15429}

\bibitem[{{Caldiroli, Andrea} {et~al.}(2021){Caldiroli, Andrea}, {Haardt, Francesco}, {Gallo, Elena}, {Spinelli, Riccardo}, {Malsky, Isaac}, \& {Rauscher, Emily}}]{Caldiroli21_ATES}
{Caldiroli, Andrea}, {Haardt, Francesco}, {Gallo, Elena}, {et~al.} 2021, A\&A, 655, A30, \dodoi{10.1051/0004-6361/202141497}

\bibitem[{Carolan {et~al.}(2021)Carolan, Vidotto, Hazra, Villarreal D’Angelo, \& Kubyshkina}]{Carolan21_obsmagnetic}
Carolan, S., Vidotto, A.~A., Hazra, G., Villarreal D’Angelo, C., \& Kubyshkina, D. 2021, Monthly Notices of the Royal Astronomical Society, 508, 6001, \dodoi{10.1093/mnras/stab2947}

\bibitem[{{Casasayas-Barris, N.} {et~al.}(2018){Casasayas-Barris, N.}, {Pallé, E.}, {Yan, F.}, {Chen, G.}, {Albrecht, S.}, {Nortmann, L.}, {Van Eylen, V.}, {Snellen, I.}, {Talens, G. J. J.}, {González Hernández, J. I.}, {Rebolo, R.}, \& {Otten, G. P. P. L.}}]{CasasayasBarris18_kelthalpha}
{Casasayas-Barris, N.}, {Pallé, E.}, {Yan, F.}, {et~al.} 2018, A\&A, 616, A151, \dodoi{10.1051/0004-6361/201832963}

\bibitem[{{Chatzikos} {et~al.}(2023){Chatzikos}, {Bianchi}, {Camilloni}, {Chakraborty}, {Gunasekera}, {Guzm{\'a}n}, {Milby}, {Sarkar}, {Shaw}, {van Hoof}, \& {Ferland}}]{Chatzikos23_CL23}
{Chatzikos}, M., {Bianchi}, S., {Camilloni}, F., {et~al.} 2023, \rmxaa, 59, 327, \dodoi{10.22201/ia.01851101p.2023.59.02.12}

\bibitem[{Christie {et~al.}(2013)Christie, Arras, \& Li}]{Christie13_Halpha}
Christie, D., Arras, P., \& Li, Z.-Y. 2013, The Astrophysical Journal, 772, 144, \dodoi{10.1088/0004-637x/772/2/144}

\bibitem[{{Czesla, S.} {et~al.}(2022){Czesla, S.}, {Lampón, M.}, {Sanz-Forcada, J.}, {García Muñoz, A.}, {López-Puertas, M.}, {Nortmann, L.}, {Yan, D.}, {Nagel, E.}, {Yan, F.}, {Schmitt, J. H. M. M.}, {Aceituno, J.}, {Amado, P. J.}, {Caballero, J. A.}, {Casasayas-Barris, N.}, {Henning, Th.}, {Khalafinejad, S.}, {Molaverdikhani, K.}, {Montes, D.}, {Pallé, E.}, {Reiners, A.}, {Schneider, P. C.}, {Ribas, I.}, {Quirrenbach, A.}, {Zapatero Osorio, M. R.}, \& {Zechmeister, M.}}]{Czesla22_hatp32}
{Czesla, S.}, {Lampón, M.}, {Sanz-Forcada, J.}, {et~al.} 2022, A\&A, 657, A6, \dodoi{10.1051/0004-6361/202039919}

\bibitem[{{Dos Santos} {et~al.}(2023){Dos Santos}, {Garc{\'\i}a Mu{\~n}oz}, {Sing}, {L{\'o}pez-Morales}, {Alam}, {Bourrier}, {Ehrenreich}, {Henry}, {Lecavelier des Etangs}, {Mikal-Evans}, {Nikolov}, {Sanz-Forcada}, \& {Wakeford}}]{DosSantos23_hd189733}
{Dos Santos}, L.~A., {Garc{\'\i}a Mu{\~n}oz}, A., {Sing}, D.~K., {et~al.} 2023, \aj, 166, 89, \dodoi{10.3847/1538-3881/ace445}

\bibitem[{{dos Santos, Leonardo A.} {et~al.}(2020){dos Santos, Leonardo A.}, {Ehrenreich, David}, {Bourrier, Vincent}, {Astudillo-Defru, Nicola}, {Bonfils, Xavier}, {Forget, François}, {Lovis, Christophe}, {Pepe, Francesco}, \& {Udry, Stéphane}}]{DosSantos20_k218}
{dos Santos, Leonardo A.}, {Ehrenreich, David}, {Bourrier, Vincent}, {et~al.} 2020, A\&A, 634, L4, \dodoi{10.1051/0004-6361/201937327}

\bibitem[{{Dos Santos, Leonardo A.} {et~al.}(2022){Dos Santos, Leonardo A.}, {Vidotto, Aline A.}, {Vissapragada, Shreyas}, {Alam, Munazza K.}, {Allart, Romain}, {Bourrier, Vincent}, {Kirk, James}, {Seidel, Julia V.}, \& {Ehrenreich, David}}]{DosSantos22_pwinds}
{Dos Santos, Leonardo A.}, {Vidotto, Aline A.}, {Vissapragada, Shreyas}, {et~al.} 2022, A\&A, 659, A62, \dodoi{10.1051/0004-6361/202142038}

\bibitem[{{Draine}(2011)}]{Draine11_IGM}
{Draine}, B.~T. 2011, {Physics of the Interstellar and Intergalactic Medium}

\bibitem[{{Ehrenreich} {et~al.}(2015){Ehrenreich}, {Bourrier}, {Wheatley}, {Lecavelier des Etangs}, {H{\'e}brard}, {Udry}, {Bonfils}, {Delfosse}, {D{\'e}sert}, {Sing}, \& {Vidal-Madjar}}]{Ehrenreich15_gj436}
{Ehrenreich}, D., {Bourrier}, V., {Wheatley}, P.~J., {et~al.} 2015, \nat, 522, 459, \dodoi{10.1038/nature14501}

\bibitem[{{Ferland} {et~al.}(2013){Ferland}, {Porter}, {van Hoof}, {Williams}, {Abel}, {Lykins}, {Shaw}, {Henney}, \& {Stancil}}]{ferland_2013}
{Ferland}, G.~J., {Porter}, R.~L., {van Hoof}, P.~A.~M., {et~al.} 2013, \rmxaa, 49, 137, \dodoi{10.48550/arXiv.1302.4485}

\bibitem[{{France} {et~al.}(2022){France}, {Fleming}, {Youngblood}, {Mason}, {Drake}, {Amerstorfer}, {Barstow}, {Bourrier}, {Champey}, {Fossati}, {Froning}, {Green}, {Gris{\'e}}, {Gronoff}, {Hellickson}, {Jin}, {Koskinen}, {Kowalski}, {Kruczek}, {Linsky}, {Lipscy}, {McEntaffer}, {McKenzie}, {Miles}, {Patton}, {Savage}, {Siegmund}, {Spittler}, {Unruh}, \& {Volz}}]{France22_EUV}
{France}, K., {Fleming}, B., {Youngblood}, A., {et~al.} 2022, Journal of Astronomical Telescopes, Instruments, and Systems, 8, 014006, \dodoi{10.1117/1.JATIS.8.1.014006}

\bibitem[{Fu {et~al.}(2024)Fu, Welbanks, Deming, Inglis, Zhang, Lothringer, Ih, Moses, Schlawin, Knutson, Henry, Greene, Sing, Savel, Kempton, Louie, Line, \& Nixon}]{Fu2024_hd189733}
Fu, G., Welbanks, L., Deming, D., {et~al.} 2024, Nature, 632, 752, \dodoi{10.1038/s41586-024-07760-y}

\bibitem[{{Fulton} \& {Petigura}(2018)}]{Fulton18_keplervii}
{Fulton}, B.~J., \& {Petigura}, E.~A. 2018, \aj, 156, 264, \dodoi{10.3847/1538-3881/aae828}

\bibitem[{Fulton {et~al.}(2017)Fulton, Petigura, Howard, Isaacson, Marcy, Cargile, Hebb, Weiss, Johnson, Morton, Sinukoff, Crossfield, \& Hirsch}]{Fulton17_radiusgap}
Fulton, B.~J., Petigura, E.~A., Howard, A.~W., {et~al.} 2017, The Astronomical Journal, 154, 109, \dodoi{10.3847/1538-3881/aa80eb}

\bibitem[{Gibson {et~al.}(2024)Gibson, Howard, Rider, Halverson, Roy, Baker, Edelstein, Smith, Fulton, Walawender, Brodheim, Brown, Chan, Dai, Deich, Gottschalk, Grillo, Hale, Hill, Holden, Householder, Isaacson, Ishikawa, Jelinsky, Kassis, Kaye, Laher, Lanclos, Lee, Lilley, McCarney, Miller, Payne, Petigura, Poppett, Raffanti, Rubenzahl, Sandford, Schwab, Shaum, Sirk, Smith, Thorne, Valliant, Vandenberg, Wang, Wishnow, Wold, Yeh, Baca, Beichman, Berriman, Brown, Casey, Chin, Chong, Cowley, Devenot, Elwir, Finstad, Fraysse, James, Jhoti, Killian, Levine, Li, Marin, Milner, Nance, O'Hanlon, Orr, Ortiz-Soto, Payne, Pember, Raskin, Savage, Seifahrt, Smith, Storesund, St{\"u}rmer, Suominen, Tehero, Boeckmann, Wages, Weisfeiler, Wilcox, Wizinowich, \& Wolfenberger}]{Gibson24_KPF}
Gibson, S.~R., Howard, A.~W., Rider, K., {et~al.} 2024, in Ground-based and Airborne Instrumentation for Astronomy X, ed. J.~J. Bryant, K.~Motohara, \& J.~R.~D. Vernet, Vol. 13096, International Society for Optics and Photonics (SPIE), 1309609, \dodoi{10.1117/12.3017841}

\bibitem[{{Guilluy} {et~al.}(2024){Guilluy}, {D'Arpa}, {Bonomo}, {Spinelli}, {Biassoni}, {Fossati}, {Maggio}, {Giacobbe}, {Lanza}, {Sozzetti}, {Borsa}, {Rainer}, {Micela}, {Affer}, {Andreuzzi}, {Bignamini}, {Boschin}, {Carleo}, {Cecconi}, {Desidera}, {Fardella}, {Ghedina}, {Mantovan}, {Mancini}, {Nascimbeni}, {Knapic}, {Pedani}, {Petralia}, {Pino}, {Scandariato}, {Sicilia}, {Stangret}, \& {Zingales}}]{Guilluy24_GAPS_He}
{Guilluy}, G., {D'Arpa}, M.~C., {Bonomo}, A.~S., {et~al.} 2024, \aap, 686, A83, \dodoi{10.1051/0004-6361/202348997}

\bibitem[{Hotnisky {et~al.}(2024)Hotnisky, Kanodia, Libby-Roberts, Mahadevan, Canas, Gupta, Han, Kobulnicky, Larsen, Robertson, Rodruck, Stefansson, Cochran, Delamer, Diddams, Fernandes, Halverson, Hebb, Lin, Monson, Ninan, Roy, \& Schwab}]{Hotnisky24_GEMS}
Hotnisky, A., Kanodia, S., Libby-Roberts, J., {et~al.} 2024, Searching for GEMS: Two Super-Jupiters around M-dwarfs -- Signatures of Instability or Accretion?
\newblock \doarXiv{2411.08159}

\bibitem[{Husser {et~al.}(2013)Husser, Wende-von Berg, Dreizler, Homeier, Reiners, Barman, \& Hauschildt}]{Husser13_PHOENIX}
Husser, T.-O., Wende-von Berg, S., Dreizler, S., {et~al.} 2013, Astronomy \& Astrophysics, 553, A6, \dodoi{10.1051/0004-6361/201219058}

\bibitem[{Jensen {et~al.}(2012)Jensen, Redfield, Endl, Cochran, Koesterke, \& Barman}]{Jensen12_halpha}
Jensen, A.~G., Redfield, S., Endl, M., {et~al.} 2012, The Astrophysical Journal, 751, 86, \dodoi{10.1088/0004-637X/751/2/86}

\bibitem[{{Johnstone} {et~al.}(2018){Johnstone}, {G{\"u}del}, {Lammer}, \& {Kislyakova}}]{johnstone_2018}
{Johnstone}, C.~P., {G{\"u}del}, M., {Lammer}, H., \& {Kislyakova}, K.~G. 2018, \aap, 617, A107, \dodoi{10.1051/0004-6361/201832776}

\bibitem[{{Kanodia} {et~al.}(2024){Kanodia}, {Gupta}, {Ca{\~n}as}, {Bernab{\`o}}, {Reji}, {Han}, {Brady}, {Seifahrt}, {Cochran}, {Morrell}, {Basant}, {Bean}, {Bender}, {de Beurs}, {Bieryla}, {Birkholz}, {Brown}, {Chapman}, {Ciardi}, {Clark}, {Cotter}, {Diddams}, {Halverson}, {Hawley}, {Hebb}, {Holcomb}, {Howell}, {Kobulnicky}, {Kowalski}, {Larsen}, {Libby-Roberts}, {Lin}, {Lund}, {Luque}, {Monson}, {Ninan}, {Parker}, {Patel}, {Rodruck}, {Ross}, {Roy}, {Schwab}, {Stef{\'a}nsson}, {Thoms}, \& {Vanderburg}}]{kanodia_2024}
{Kanodia}, S., {Gupta}, A.~F., {Ca{\~n}as}, C.~I., {et~al.} 2024, \aj, 168, 235, \dodoi{10.3847/1538-3881/ad7796}

\bibitem[{{Kirk} {et~al.}(2020){Kirk}, {Alam}, {L{\'o}pez-Morales}, \& {Zeng}}]{kirk_2020}
{Kirk}, J., {Alam}, M.~K., {L{\'o}pez-Morales}, M., \& {Zeng}, L. 2020, \aj, 159, 115, \dodoi{10.3847/1538-3881/ab6e66}

\bibitem[{Kubyshkina {et~al.}(2023)Kubyshkina, Fossati, \& Erkaev}]{kubyshkina_2023}
Kubyshkina, D., Fossati, L., \& Erkaev, N.~V. 2023, Precise photoionisation treatment and hydrodynamic effects in atmospheric modelling of warm and hot Neptunes.
\newblock \doarXiv{2312.07236}

\bibitem[{{Lamp{\'o}n} {et~al.}(2021){Lamp{\'o}n}, {L{\'o}pez-Puertas}, {Czesla}, {S{\'a}nchez-L{\'o}pez}, {Lara}, {Salz}, {Sanz-Forcada}, {Molaverdikhani}, {Quirrenbach}, {Pall{\'e}}, {Caballero}, {Henning}, {Nortmann}, {Amado}, {Montes}, {Reiners}, \& {Ribas}}]{Lampon21_escobs}
{Lamp{\'o}n}, M., {L{\'o}pez-Puertas}, M., {Czesla}, S., {et~al.} 2021, \aap, 648, L7, \dodoi{10.1051/0004-6361/202140423}

\bibitem[{Lampón {et~al.}(2021)Lampón, López-Puertas, Sanz-Forcada, Sánchez-López, Molaverdikhani, Czesla, Quirrenbach, Pallé, Caballero, Henning, Salz, Nortmann, Aceituno, Amado, Bauer, Montes, Nagel, Reiners, \& Ribas}]{Lampon21_hd189733}
Lampón, M., López-Puertas, M., Sanz-Forcada, J., {et~al.} 2021, Astronomy \& Astrophysics, 647, A129, \dodoi{10.1051/0004-6361/202039417}

\bibitem[{{Lampón, M.} {et~al.}(2020){Lampón, M.}, {López-Puertas, M.}, {Lara, L. M.}, {Sánchez-López, A.}, {Salz, M.}, {Czesla, S.}, {Sanz-Forcada, J.}, {Molaverdikhani, K.}, {Alonso-Floriano, F. J.}, {Nortmann, L.}, {Caballero, J. A.}, {Bauer, F. F.}, {Pallé, E.}, {Montes, D.}, {Quirrenbach, A.}, {Nagel, E.}, {Ribas, I.}, {Reiners, A.}, \& {Amado, P. J.}}]{Lampon20_hd209458}
{Lampón, M.}, {López-Puertas, M.}, {Lara, L. M.}, {et~al.} 2020, A\&A, 636, A13, \dodoi{10.1051/0004-6361/201937175}

\bibitem[{{Lampón, M.} {et~al.}(2023){Lampón, M.}, {López-Puertas, M.}, {Sanz-Forcada, J.}, {Czesla, S.}, {Nortmann, L.}, {Casasayas-Barris, N.}, {Orell-Miquel, J.}, {Sánchez-López, A.}, {Danielski, C.}, {Pallé, E.}, {Molaverdikhani, K.}, {Henning, Th.}, {Caballero, J. A.}, {Amado, P. J.}, {Quirrenbach, A.}, {Reiners, A.}, \& {Ribas, I.}}]{Lampon23_he}
{Lampón, M.}, {López-Puertas, M.}, {Sanz-Forcada, J.}, {et~al.} 2023, A\&A, 673, A140, \dodoi{10.1051/0004-6361/202245649}

\bibitem[{{Lecavelier Des Etangs} {et~al.}(2010){Lecavelier Des Etangs}, {Ehrenreich}, {Vidal-Madjar}, {Ballester}, {D{\'e}sert}, {Ferlet}, {H{\'e}brard}, {Sing}, {Tchakoumegni}, \& {Udry}}]{Lecavelier10_hd189}
{Lecavelier Des Etangs}, A., {Ehrenreich}, D., {Vidal-Madjar}, A., {et~al.} 2010, \aap, 514, A72, \dodoi{10.1051/0004-6361/200913347}

\bibitem[{{Linssen} {et~al.}(2024){Linssen}, {Shih}, {MacLeod}, \& {Oklop{\v{c}}i{\'c}}}]{Linssen24_sunbather}
{Linssen}, D., {Shih}, J., {MacLeod}, M., \& {Oklop{\v{c}}i{\'c}}, A. 2024, arXiv e-prints, arXiv:2404.12775, \dodoi{10.48550/arXiv.2404.12775}

\bibitem[{{Linssen, D. C.} {et~al.}(2022){Linssen, D. C.}, {Oklopčić, A.}, \& {MacLeod, M.}}]{Linssen22_cloudy}
{Linssen, D. C.}, {Oklopčić, A.}, \& {MacLeod, M.} 2022, A\&A, 667, A54, \dodoi{10.1051/0004-6361/202243830}

\bibitem[{{McLean} {et~al.}(1998){McLean}, {Becklin}, {Bendiksen}, {Brims}, {Canfield}, {Figer}, {Graham}, {Hare}, {Lacayanga}, {Larkin}, {Larson}, {Levenson}, {Magnone}, {Teplitz}, \& {Wong}}]{McLean98_NIRSPEC}
{McLean}, I.~S., {Becklin}, E.~E., {Bendiksen}, O., {et~al.} 1998, in Society of Photo-Optical Instrumentation Engineers (SPIE) Conference Series, Vol. 3354, Infrared Astronomical Instrumentation, ed. A.~M. {Fowler}, 566--578, \dodoi{10.1117/12.317283}

\bibitem[{{Mignone} {et~al.}(2007){Mignone}, {Bodo}, {Massaglia}, {Matsakos}, {Tesileanu}, {Zanni}, \& {Ferrari}}]{Mignone07_PLUTO}
{Mignone}, A., {Bodo}, G., {Massaglia}, S., {et~al.} 2007, \apjs, 170, 228, \dodoi{10.1086/513316}

\bibitem[{{Mo{\v{c}}nik} {et~al.}(2017){Mo{\v{c}}nik}, {Hellier}, {Anderson}, {Clark}, \& {Southworth}}]{Mocnik17_wasp107}
{Mo{\v{c}}nik}, T., {Hellier}, C., {Anderson}, D.~R., {Clark}, B.~J.~M., \& {Southworth}, J. 2017, \mnras, 469, 1622, \dodoi{10.1093/mnras/stx972}

\bibitem[{{Murray-Clay} {et~al.}(2009){Murray-Clay}, {Chiang}, \& {Murray}}]{MurrayClay09_escape}
{Murray-Clay}, R.~A., {Chiang}, E.~I., \& {Murray}, N. 2009, \apj, 693, 23, \dodoi{10.1088/0004-637X/693/1/23}

\bibitem[{Nortmann {et~al.}(2018)Nortmann, Pallé, Salz, Sanz-Forcada, Nagel, Alonso-Floriano, Czesla, Yan, Chen, Snellen, Zechmeister, Schmitt, López-Puertas, Casasayas-Barris, Bauer, Amado, Caballero, Dreizler, Henning, Lampón, Montes, Molaverdikhani, Quirrenbach, Reiners, Ribas, Sánchez-López, Schneider, \& Osorio}]{Nortmann18_wasp69}
Nortmann, L., Pallé, E., Salz, M., {et~al.} 2018, Science, 362, 1388, \dodoi{10.1126/science.aat5348}

\bibitem[{{Oklop{\v{c}}i{\'c}}(2019)}]{Oklopcic19_He}
{Oklop{\v{c}}i{\'c}}, A. 2019, \apj, 881, 133, \dodoi{10.3847/1538-4357/ab2f7f}

\bibitem[{{Oklop{\v{c}}i{\'c}} \& {Hirata}(2018)}]{Oklopcic18_He}
{Oklop{\v{c}}i{\'c}}, A., \& {Hirata}, C.~M. 2018, \apjl, 855, L11, \dodoi{10.3847/2041-8213/aaada9}

\bibitem[{{Orell-Miquel} {et~al.}(2024){Orell-Miquel}, {Murgas}, {Pall{\'e}}, {Mallorqu{\'\i}n}, {L{\'o}pez-Puertas}, {Lamp{\'o}n}, {Sanz-Forcada}, {Nortmann}, {Czesla}, {Nagel}, {Ribas}, {Stangret}, {Livingston}, {Knudstrup}, {Albrecht}, {Carleo}, {Caballero}, {Dai}, {Esparza-Borges}, {Fukui}, {Heng}, {Henning}, {Kagetani}, {Lesjak}, {de Leon}, {Montes}, {Morello}, {Narita}, {Quirrenbach}, {Amado}, {Reiners}, {Schweitzer}, \& {Vico Linares}}]{OrellMiquel24_MOPYS_preprint}
{Orell-Miquel}, J., {Murgas}, F., {Pall{\'e}}, E., {et~al.} 2024, arXiv e-prints, arXiv:2404.16732, \dodoi{10.48550/arXiv.2404.16732}

\bibitem[{Owen \& Adams(2019)}]{Owen19_valleymag}
Owen, J.~E., \& Adams, F.~C. 2019, Monthly Notices of the Royal Astronomical Society, 490, 15, \dodoi{10.1093/mnras/stz2601}

\bibitem[{Owen \& Schlichting(2023)}]{Owen23_cpmassloss}
Owen, J.~E., \& Schlichting, H.~E. 2023, Mapping out the parameter space for photoevaporation and core-powered mass-loss.
\newblock \doarXiv{2308.00020}

\bibitem[{{Owen} \& {Wu}(2013)}]{Owen13_kepler}
{Owen}, J.~E., \& {Wu}, Y. 2013, \apj, 775, 105, \dodoi{10.1088/0004-637X/775/2/105}

\bibitem[{Salz {et~al.}(2015)Salz, Banerjee, Mignone, Schneider, Czesla, \& Schmitt}]{Salz15_TPCI}
Salz, M., Banerjee, R., Mignone, A., {et~al.} 2015, Astronomy \& Astrophysics, 576, A21, \dodoi{10.1051/0004-6361/201424330}

\bibitem[{{Salz} {et~al.}(2016){Salz}, {Czesla}, {Schneider}, \& {Schmitt}}]{Salz16_escaping}
{Salz}, M., {Czesla}, S., {Schneider}, P.~C., \& {Schmitt}, J.~H.~M.~M. 2016, \aap, 586, A75, \dodoi{10.1051/0004-6361/201526109}

\bibitem[{{Salz, M.} {et~al.}(2018){Salz, M.}, {Czesla, S.}, {Schneider, P. C.}, {Nagel, E.}, {Schmitt, J. H. M. M.}, {Nortmann, L.}, {Alonso-Floriano, F. J.}, {López-Puertas, M.}, {Lampón, M.}, {Bauer, F. F.}, {Snellen, I. A. G.}, {Pallé, E.}, {Caballero, J. A.}, {Yan, F.}, {Chen, G.}, {Sanz-Forcada, J.}, {Amado, P. J.}, {Quirrenbach, A.}, {Ribas, I.}, {Reiners, A.}, {Béjar, V. J. S.}, {Casasayas-Barris, N.}, {Cortés-Contreras, M.}, {Dreizler, S.}, {Guenther, E. W.}, {Henning, T.}, {Jeffers, S. V.}, {Kaminski, A.}, {Kürster, M.}, {Lafarga, M.}, {Lara, L. M.}, {Molaverdikhani, K.}, {Montes, D.}, {Morales, J. C.}, {Sánchez-López, A.}, {Seifert, W.}, {Zapatero Osorio, M. R.}, \& {Zechmeister, M.}}]{Salz18_hd189733}
{Salz, M.}, {Czesla, S.}, {Schneider, P. C.}, {et~al.} 2018, A\&A, 620, A97, \dodoi{10.1051/0004-6361/201833694}

\bibitem[{Sanz-Forcada {et~al.}(2025)Sanz-Forcada, Lopez-Puertas, Lamp’on, Czesla, Nortmann, Caballero, Zapatero~Osorio, Amado, Murgas, Orell-Miquel, Pall’e, Quirrenbach, Reiners, Ribas, Sanchez-Lopez, \& Solano}]{SanzForcada25_He}
Sanz-Forcada, J., Lopez-Puertas, M., Lamp’on, M., {et~al.} 2025, Astronomy \& Astrophysics, \dodoi{10.1051/0004-6361/202451680}

\bibitem[{Schreyer {et~al.}(2023)Schreyer, Owen, Spake, Bahroloom, \& Di Giampasquale}]{Schreyer24_mag}
Schreyer, E., Owen, J.~E., Spake, J.~J., Bahroloom, Z., \& Di Giampasquale, S. 2023, Monthly Notices of the Royal Astronomical Society, 527, 5117, \dodoi{10.1093/mnras/stad3528}

\bibitem[{{Seager} \& {Sasselov}(2000)}]{Seager00_hd209458}
{Seager}, S., \& {Sasselov}, D.~D. 2000, \apj, 537, 916, \dodoi{10.1086/309088}

\bibitem[{Spake {et~al.}(2018)Spake, Sing, Evans, Oklop{\v{c}}i{\'{c}}, Bourrier, Kreidberg, Rackham, Irwin, Ehrenreich, Wyttenbach, Wakeford, Zhou, Chubb, Nikolov, Goyal, Henry, Williamson, Blumenthal, Anderson, Hellier, Charbonneau, Udry, \& Madhusudhan}]{Spake2018_wasp107}
Spake, J.~J., Sing, D.~K., Evans, T.~M., {et~al.} 2018, Nature, 557, 68, \dodoi{10.1038/s41586-018-0067-5}

\bibitem[{{Stassun} {et~al.}(2017){Stassun}, {Collins}, \& {Gaudi}}]{Stassun17_wasp69}
{Stassun}, K.~G., {Collins}, K.~A., \& {Gaudi}, B.~S. 2017, \aj, 153, 136, \dodoi{10.3847/1538-3881/aa5df3}

\bibitem[{{Vidal-Madjar} {et~al.}(2003){Vidal-Madjar}, {Lecavelier des Etangs}, {D{\'e}sert}, {Ballester}, {Ferlet}, {H{\'e}brard}, \& {Mayor}}]{VidalMadjar03_hd209}
{Vidal-Madjar}, A., {Lecavelier des Etangs}, A., {D{\'e}sert}, J.~M., {et~al.} 2003, \nat, 422, 143, \dodoi{10.1038/nature01448}

\bibitem[{{Wang} \& {Dai}(2021)}]{Wang21_wasp69}
{Wang}, L., \& {Dai}, F. 2021, \apj, 914, 98, \dodoi{10.3847/1538-4357/abf1ee}

\bibitem[{{Wang} {et~al.}(2019){Wang}, {Wang}, {Hinse}, {Wu}, {Davis}, {Hori}, {Yoon}, {Han}, {Nie}, {Liu}, {Zhang}, {Zhou}, {Wittenmyer}, {Peng}, \& {Laughlin}}]{Wang19_hatp32}
{Wang}, Y.-H., {Wang}, S., {Hinse}, T.~C., {et~al.} 2019, \aj, 157, 82, \dodoi{10.3847/1538-3881/aaf6b6}

\bibitem[{{Wolfgang} {et~al.}(2016){Wolfgang}, {Rogers}, \& {Ford}}]{Wolfgang16_subneptune}
{Wolfgang}, A., {Rogers}, L.~A., \& {Ford}, E.~B. 2016, \apj, 825, 19, \dodoi{10.3847/0004-637X/825/1/19}

\bibitem[{Xue {et~al.}(2024)Xue, Bean, Zhang, Welbanks, Lunine, \& August}]{Xue24_hd209458}
Xue, Q., Bean, J.~L., Zhang, M., {et~al.} 2024, The Astrophysical Journal Letters, 963, L5, \dodoi{10.3847/2041-8213/ad2682}

\bibitem[{Yoshida {et~al.}(2024)Yoshida, Terada, \& Kuramoto}]{Yoshida24_molecules}
Yoshida, T., Terada, N., \& Kuramoto, K. 2024, Progress in Earth and Planetary Science, 11, \dodoi{10.1186/s40645-024-00666-3}

\bibitem[{Zhang {et~al.}(2022{\natexlab{a}})Zhang, Cauley, Knutson, France, Kreidberg, Oklopčić, Redfield, \& Shkolnik}]{Zhang22_hd189733}
Zhang, M., Cauley, P.~W., Knutson, H.~A., {et~al.} 2022{\natexlab{a}}, The Astronomical Journal, 164, 237, \dodoi{10.3847/1538-3881/ac9675}

\bibitem[{{Zhang} {et~al.}(2023){Zhang}, {Dai}, {Bean}, {Knutson}, \& {Rescigno}}]{Zhang23_toi2134}
{Zhang}, M., {Dai}, F., {Bean}, J.~L., {Knutson}, H.~A., \& {Rescigno}, F. 2023, \apjl, 953, L25, \dodoi{10.3847/2041-8213/aced51}

\bibitem[{Zhang {et~al.}(2023)Zhang, Knutson, Dai, Wang, Ricker, Schwarz, Mann, \& Collins}]{Zhang23_neptunes}
Zhang, M., Knutson, H.~A., Dai, F., {et~al.} 2023, The Astronomical Journal, 165, 62, \dodoi{10.3847/1538-3881/aca75b}

\bibitem[{Zhang {et~al.}(2022{\natexlab{b}})Zhang, Knutson, Wang, Dai, \& Barragán}]{Zhang22_toi560}
Zhang, M., Knutson, H.~A., Wang, L., Dai, F., \& Barragán, O. 2022{\natexlab{b}}, The Astronomical Journal, 163, 67, \dodoi{10.3847/1538-3881/ac3fa7}

\bibitem[{Zhang \& Rosener(2024)}]{Zhang_Rosener_2024}
Zhang, M., \& Rosener, R. 2024, pyTPCI (Zenodo), \dodoi{10.5281/ZENODO.14285142}

\bibitem[{Zhang {et~al.}(2022{\natexlab{c}})Zhang, Knutson, Wang, Dai, dos Santos, Fossati, Henry, Ehrenreich, Alibert, Hoyer, Wilson, \& Bonfanti}]{Zhang22_hd63433}
Zhang, M., Knutson, H.~A., Wang, L., {et~al.} 2022{\natexlab{c}}, The Astronomical Journal, 163, 68, \dodoi{10.3847/1538-3881/ac3f3b}

\bibitem[{Zhang {et~al.}(2024)Zhang, Bean, Wilson, Duvvuri, Schneider, Knutson, Dai, Collins, Watkins, Schwarz, Barkaoui, Shporer, Horne, Sefako, Murgas, \& Palle}]{Zhang24_toi836}
Zhang, M., Bean, J.~L., Wilson, D., {et~al.} 2024, Constraining atmospheric composition from the outflow: helium observations reveal the fundamental properties of two planets straddling the radius gap.
\newblock \doarXiv{2409.08318}

\bibitem[{{Zhang} {et~al.}(2024){Zhang}, {Bean}, {Wilson}, {Duvvuri}, {Schneider}, {Knutson}, {Dai}, {Collins}, {Watkins}, {Schwarz}, {Barkaoui}, {Shporer}, {Horne}, {Sefako}, {Murgas}, \& {Palle}}]{zhang_2024}
{Zhang}, M., {Bean}, J.~L., {Wilson}, D., {et~al.} 2024, arXiv e-prints, arXiv:2409.08318, \dodoi{10.48550/arXiv.2409.08318}

\bibitem[{{Zhang} {et~al.}(2020){Zhang}, {Snellen}, {Molli{\`e}re}, {Alonso-Floriano}, {Webb}, {Brogi}, \& {Wyttenbach}}]{ZhangSnellen20_He}
{Zhang}, Y., {Snellen}, I.~A.~G., {Molli{\`e}re}, P., {et~al.} 2020, \aap, 641, A161, \dodoi{10.1051/0004-6361/202038412}

\end{thebibliography}
\bibliographystyle{aasjournal}



\end{document}